\documentclass[reprint,
 amsmath,amssymb,
 aps,
showkeys]{revtex4-2}

\usepackage{graphicx}
\usepackage{dcolumn}
\usepackage{bm}
\usepackage{physics}
\usepackage{cleveref}
\usepackage{xcolor}
\usepackage{bbold}
\allowdisplaybreaks

\DeclareUnicodeCharacter{2212}{-}

\begin{document}


\title{Electron-assisted manipulation of polaritonic light-matter states}

\author{J. Abad-Arredondo}
\email{jaime.abad(at)uam.es}
\author{A. I. Fern\'andez-Dom\'inguez}%

\affiliation{Departamento de F\'isica Te\'orica de la Materia
Condensada and Condensed Matter Physics Center (IFIMAC),
Universidad Aut\'onoma de Madrid, E- 28049 Madrid, Spain 
}%

\date{\today}

\begin{abstract}
Thanks to their exceptional spatial, spectral and temporal resolution, highly-coherent free-electron beams have emerged as powerful probes for material excitations, enabling their characterization even in the quantum regime. Here, we investigate strong light-matter coupling through monochromatic and modulated electron wavepackets. In particular, we consider an archetypal target, comprising a nanophotonic cavity next to a single two-level emitter.  We propose a model Hamiltonian describing the coherent interaction between the passing electron beam and the hybrid photonic-excitonic target, which is constructed using macroscopic quantum electrodynamics and fully parameterized in terms of the electromagnetic Dyadic Green’s function. Using this framework, we first describe electron-energy-loss and cathodoluminescence spectroscopies, and photon-induced near-field electron emission microscopy. Finally, we show the power of modulated electrons beams as quantum tools for the manipulation of polaritonic targets presenting a complex energy landscape of excitations.
\end{abstract}


\maketitle

\section{Introduction}

Much research attention has focused lately on the strong-coupling (SC) phenomena that emerge when quantum emitters (QEs), such as organic molecules, solid-state vacancies, or quantum dots, are placed within the near-field of photonic resonators, such as Fabry-Perot cavities, metamaterial devices, or nanoantennas~\cite{Novotny2010,Torma2014,Tserkezis2020}. In setups involving macroscopic ensembles of QEs, the formation of polaritons (hybrid light-matter states) has opened the way for the manipulation of matter for purposes such as the modification of material properties or the control of chemical reactions~\cite{Sanvitto2016,GarciaVidal2021}. The high complexity of these systems, however, makes their theoretical description extremely challenging, which severely limits the capability of current theories to reproduce experimental results~\cite{Tserkezis2020,SanchezBarquilla2022}. Complementarily, polariton formation in systems comprising a single (or few) QEs~\cite{Lodahl2015,Santhosh2016,Chikkaraddy2016} have been investigated for quantum light generation~\cite{Saez-Blazquez2018,FernandezDominguez2018} in studies that have also shed light into different aspects of light-matter SC at the macroscopic scale~\cite{Ojambati2019}. However, the inherent dark character of these microscopic systems~\cite{Baumberg2019}, which must feature large light-matter interaction strengths and small radiative losses, prevents their full characterization by far-field, optical means. 

Traditional electron-beam-based optical characterization methods~\cite{GarciadeAbajo2010}, such as electron-energy-loss spectroscopy (EELS) or cathodoluminiscence (CL) microscopy, present extraordinary spatial and spectral resolutions, approaching the subnanometric and milielectronvolt ranges, respectively~\cite{Duan2012,Lagos2017}. These make them ideal for the exploration of light-matter SC and polaritonic states in nanophotonic samples involving only a few excitons~\cite{Yankovich2019,Crai2020,Zouros2020,Zhu2023}. Moreover, in the last years, advances in ultrafast optical control of free-electron  wavepackets reached the femtosecond scale, matching the optical period of visible light~\cite{Polman2019}. These are behind the emergence of techniques such as photon induced near-field electron microscopy (PINEM), that exploits the synchronous interaction between free-electrons and spatially-confined pulsed laser fields~\cite{Barwick2009}. Developments in PINEM theory~\cite{GarciadeAbajo2010PINEM,Park2010} and, generally, in the description of electron-photon interactions~\cite{Talebi2018,Reinhardt2020}, together with the extraordinary degree of optical modulation (in time and momentum space) of electron beams attainable today~\cite{Priebe2017,Morimoto2018,kfir2020}, have made possible their use to imprint, exchange and manipulate quantum coherence in optical and material excitations, sustained by  micro- and nano-cavities ~\cite{Kfir2019,DiGiulio2019,Lim2022,GarciadeAbajo2021,Kfir2021} and QEs~\cite{Zhao2021,Ruimy2021,GarciadeAbajo2022}, respectively. Only very recently, similar ideas have been proposed for hybdrid excitonic-photonic systems, where light-matter SC gives rise to a much more complex ladder of polaritonic energies ~\cite{Karnieli2023}. 

Here, we present a model Hamiltonian describing the quantum interaction between a modulated electron wavepacket and a polaritonic target comprising a single QE (treated as a two-level system) and a nanophotonic cavity. The Hamiltonian is constructed using the framework of macroscopic quantum electrodynamics (QED)~\cite{Scheel2008,Dung2002,Rivera2020,Feist2020} and is fully parameterized in terms of the electromagnetic Dyadic Green’s function. For simplicity, we consider a cavity with spherical symmetry, and to unveil clearly quantum-coherent effects in the light-matter SC, we restrict its Hilbert space to the lowest (degenerate), dipolar modes that it supports. We explore the polariton energy ladder of the hybrid photonic-excitonic system through both the free-electron wavepacket and photon spectra in EELS-, CL- and PINEM-like setups. Finally, we demonstrate the power of modulated electron beams to probe and control light-matter states in the SC regime. 

\section{Target-probe system and model Hamiltonian}

\begin{figure}[!h]
\includegraphics[angle=0,width=\columnwidth]{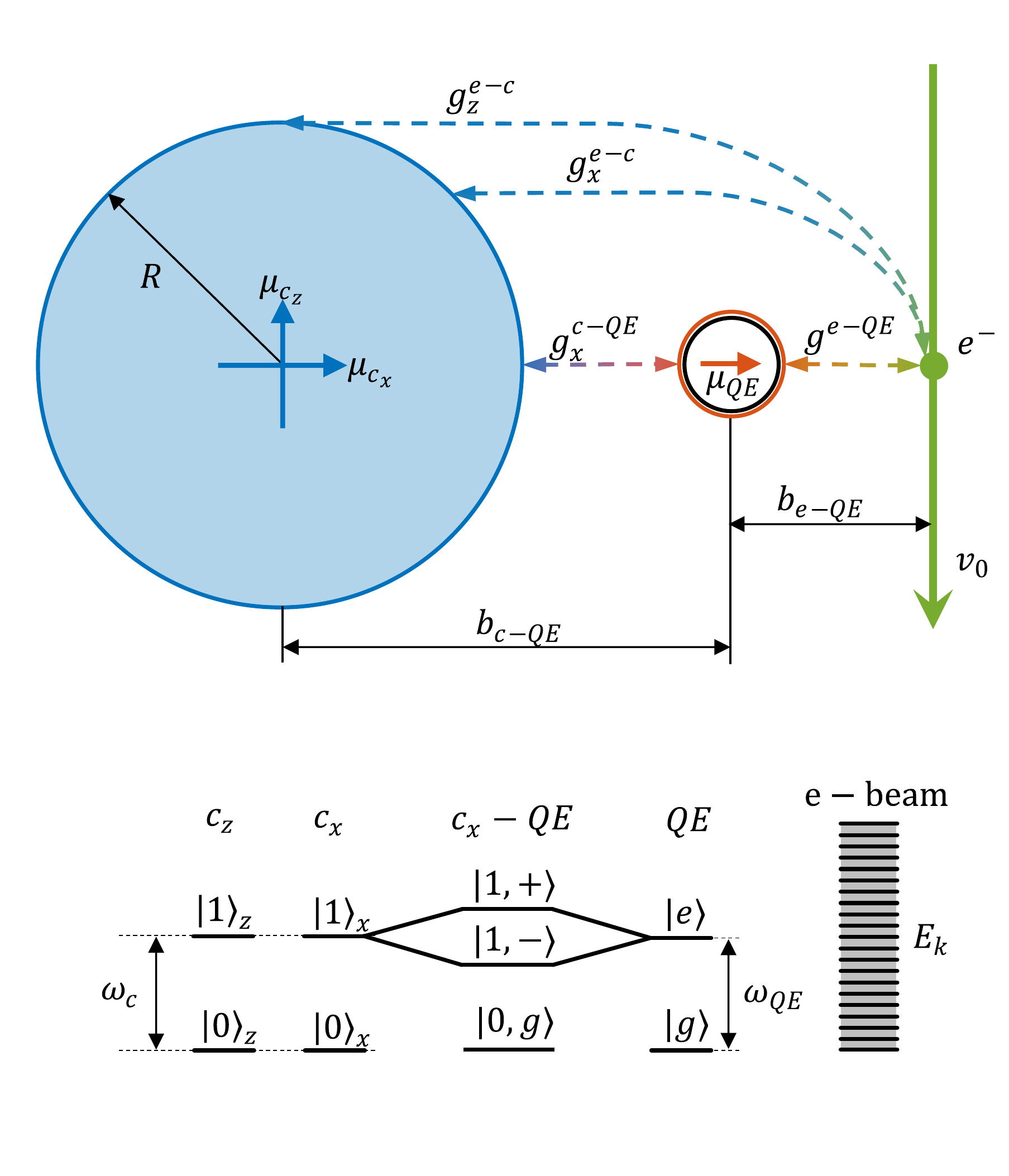}
\caption{Top: Sketch of the system under consideration: an electron wavepacket with central velocity $v_0\hat{z}$ and kinetic energy $E_{k}$ passes through a target system composed of a nanoparicle cavity and a QE. The nanoparticle radius is 10 nm and its spectrum is restricted to two degenerate dipolar cavity modes with energy $\hbar\omega_{c}=\hbar\omega_{QE}=2$ eV, at resonance with the QE. Bottom: Illustration of the energy levels of $\hat{H_0}$ for target (left) and electron beam (right). The $z$-dipolar cavity mode is uncoupled from the QE, while the $x$-dipolar one is strongly coupled to it, giving rise to non-degenerate polaritonic states.} \label{fig:fig1}
\end{figure}

The target-probe system that we have chosen to assess the ability of free electrons to explore light-matter SC is depicted in the top panel of \Cref{fig:fig1}. We consider a nanophotonic cavity (typically a metal nanoparticle), sustaining a dipolar-like confined mode overlapping with the dipole moment, $\mu_{QE}=1\,{\rm e\cdot nm}$ (parallel to $x$-axis), of a QE placed in close proximity of the nanoparticle surface (the QE-cavity distance is similar to the cavity radius itself, $b_{c-QE}\approx R$), also along the $x$-direction. The free-electron wavepacket, with energies in the order of 10 keV, passes through the compound target along the $\hat{z}$ direction with impact parameters $b_{e-c}$ and $b_{e-QE}$ with respect to cavity and QE, respectively. QE and cavity are, unless specified otherwise, at resonance, with $\hbar \omega_{c}=\hbar \omega_{QE}=2$ eV. This enables us to neglect the contribution from higher order, multipolar modes in the QE-cavity interaction. To maximize their coupling, we set $R=10$ nm, which corresponds to modal dipole moments of $\mu_{c_{x,y}}=40\, e\cdot$nm~\cite{Ballestero2015}. The small size of the cavity allows us to use the quasi-static approximation for the Dyadic Green's functions employed in the electromagnetic description of the target and passing electrons. In Sections S1-S4 of the Supplementary Material (SM), we provide details of the derivation of the system Hamiltonian and its quasi-static parametrization using macroscopic QED. It can be written as $\hat{H}= \hat{H_0}+\hat{H_I}$, with 
\begin{flalign}
\hat{H_0}=&\hbar \sum_{i=x,z}\omega_{c}\hat{a}_i^\dagger\hat{a}_i+ \hbar\omega_{QE}\hat{\sigma}^\dagger \hat{\sigma}+ \nonumber\\
&+\sum_k E_k \hat{c}_k^\dagger \hat{c}_k 
       +\hbar g^{c-QE}_x\left[\hat{a}^\dagger_x\sigma+\hat{a}_x\sigma^\dagger\right], \label{eq:simplifyied_hamiltonian1} \\
\hat{H_I}=&+\hbar \sum_{i=x,z} \sum_{q} \, g_{q,i}^{e-c}\hat{b}_q \left[\hat{a}_i^\dagger-\hat{a}_i\right] \mbox{sign}(q)\nonumber \\
    &+\hbar\sum_{q}g_{q}^{e-QE}\hat{b}_q\left[ \sigma - \sigma^{\dagger}\right]\mbox{sign}(q). \label{eq:simplifyied_hamiltonian2}
\end{flalign}
$\hat{H_0}$ describes the free dynamics of target and electron beam independently, and $\hat{H_I}$ their interaction. This Hamiltonian captures the terms previously used to study free electron interaction with optical modes~\cite{DiGiulio2019,Lim2022,GarciadeAbajo2021}, QEs~\cite{Zhao2021,Ruimy2021,GarciadeAbajo2022}, and polaritonic systems \cite{Karnieli2023}. We also note that the parametrization through macroscopic QED allows retrieving the classical results from EELS theory (see Section S4 of the SM). Thus, $\hat{a}_i$ ($i=x,z$) are the annihilation operators for the degenerate dipolar cavity modes (note that, by symmetry, we can consider only those within the $xz$-plane in \Cref{fig:fig1}), $\sigma = \ket{g}\bra{e}$ is the two-level-system lowering operator for the QE excitons, and $c_k$ is the operator describing the annihilation of free-electron population in the wavepacket component with momentum $k$ and energy $E_k=(\hbar k)^2 /2m_e$. The fourth term in Equation~\eqref{eq:simplifyied_hamiltonian1}  accounts for the cavity-emitter coupling in the rotating wave approximation with strength (see Section S4 of the SM)
 \begin{flalign}   
        &g_{x}^{c-QE}=\frac{\omega_{QE}}{3}\sqrt{\frac{\pi}{2}\left(\frac{R}{b_{c-QE}}\right)^3 \frac{\mu_{QE}^2}{\hbar \omega_{c}\epsilon_0 b_{c-QE}^3}}. \label{eq:Coupling_final_1}
\end{flalign}
Note that the QE only couples with the cavity mode with an effective dipole moment along $x$-direction. \\

The two Holstein-like terms in Equation~\eqref{eq:simplifyied_hamiltonian2} describe the target-probe interaction, where $\hat{b}_q=\sum_k\hat{c}_{k-q}^\dagger\hat{c}_k$ is the ladder operator that shifts the free-electron momentum by an amount $q$, which is transferred to or from the  cavity modes (first terms) or QE exciton (last term). Note that, contrary to the QE, the passing electrons couple to both the $x$- and $z$-dipolar cavity modes. As detailed in Sections S3-S4 of the SM, the electron-cavity and electron-QE coupling strenghts can be written as
   \begin{flalign}   
   &g_{q,x}^{e-c}=\frac{e \hbar k_0}{3m_e L } q^2K_1(|q|b_{e-c}) \sqrt{\frac{1}{ \hbar\epsilon_0} \frac{\pi}{2}\frac{ R^3}{ \omega_{c}} },\label{eq:Coupling_final_2}\\
  & g_{q,z}^{e-c}=\frac{e \hbar k_0}{3m_e L } q^2K_0(|q|b_{e-c}) \sqrt{\frac{1}{ \hbar\epsilon_0} \frac{\pi}{2}\frac{ R^3}{ \omega_{c}} },\label{eq:Coupling_final_3}\\
    & g_{q}^{e-QE}=\frac{e k_0q^2\mu_{QE}}{2\pi m_e L \epsilon_0 \omega_{QE}}
    K_1(|q| b_{e-QE}),\label{eq:Coupling_final_4}
   \end{flalign}
where $\hbar k_0=m_ev_0>>\hbar|q|$ is the incoming momentum of the passing electrons, which is $\sim 4$ orders of magnitude larger than the momentum they exchange with the cavity-emitter target ($|q|\sim\omega_{c,QE}/v_0$). This fact enables us to operate under the so-called nonrecoil approximation~\cite{GarciadeAbajo2010}. $K_{0.1}(\cdot)$ are modified Bessel functions of the second kind, and we have assumed positive impact parameters ($b_{i-j}>0$ for all $i$, $j$). $L$ is a length scale introduced in the particle-in-a-box quantization of the electron momentum, we anticipate that all the physical observables calculated in the following will not depend on this quantity, formally introduced for clarity.\\

We are interested in employing the electron beam as a tool to explore light-matter SC in the target. Therefore, we will proceed by diagonalizing (analytically) the bare Hamiltonian, $\hat{H_0}$, accounting for the cavity-QE interactions at all orders in the coupling strength $g_x^{c-QE}$ and obtaining the polaritonic eigenstates of the target. On the contrary, taking advantage of the fact that the incoming electrons only alter the target weakly, the interaction Hamiltonian, $\hat{H_I}$, will be treated perturbately, only considering processes up to second order of interaction in $g_{q,i}^{e-c}$ and $g_{q}^{e-QE}$. The bottom panel of \Cref{fig:fig1} illustrates the energy levels of the target (left) and electrons (right). Note that the energy scales are very different as, as indicated above, $E_k>>\hbar\omega_{c,QE}$. The sketch of the ground and first excitation manifolds for the target shows an uncoupled $z$-dipolar cavity mode and the emergence of polaritonic states as a result of the hybridization of the $x$-dipolar cavity mode and the QE exciton. The eigenstates of $\hat{H}_0$ can be expressed as a product of the free electron states, $\ket{k}$, the Fock states of the uncoupled cavity mode, $\ket{n}_z$, and the polaritonic states. If cavity and QE are at resonance (which is the reference configuration for our study), these can be simply written as $\ket{N,\pm}=(\ket{N}_x\ket{g}\pm\ket{N-1}_x\ket{e})/\sqrt{2}$ in the $N$-th manifold, with energies $\hbar\omega_{N,\pm}=\hbar(N\omega_{c,QE}\pm\sqrt{N}g^{c-QE}_x)$~\cite{Novotny2010,Torma2014,GarciaVidal2021}. Therefore, we have 
\begin{equation}
\hat{H_0}\ket{\phi}=E_\phi\ket{\phi}, \label{eq:phi-states}
\end{equation}
for the bare system, with $\ket{\phi}=\ket{n}_z\otimes\ket{N,\pm}\otimes\ket{k}$ and $E_\phi=\hbar\omega_{c}n_z+\hbar\omega_{N,\pm}+E_k$.

\section{Electron-target interaction}

We use the scattering matrix formalism~\cite{Ruimy2021,Kfir2021,Zhao2021} to describe the alteration of the target states by the passing electrons, which amounts to applying the propagator for the interaction Hamiltonian in the interaction picture $\hat{S}(t)=\exp\left(-\frac{i}{\hbar}\int_{-\infty}^{t}\hat{H}_{I,int}(\tau) d\tau\right)$, with $\hat{H}_{I,int}(\tau)=e^{i\hat{H}_0\tau/\hbar}\hat{H}_{I}e^{-i\hat{H}_0\tau/\hbar}$. The plasmonic nature of the cavity translates into optical mode lifetimes in the range of several tens of femtoseconds, while the QE lifetime is of the order of hundreds of ps. The electron-target interaction time can be estimated from the ratio $\lambda_c/4v_0\simeq2$ fs (where we have assumed a size for the subwavelengh-confined cavity mode of $\lambda_c/4$), which is at least one order of magnitude faster than the lifetime of the target states~\cite{Kholmicheva2018}. Thus, using the quasi-instantaneous character of the electron-target interaction, we can describe the mixing of $\ket{\phi}$ that they induce through    
\begin{flalign}
     &\hat{S}(t\rightarrow\infty)=\hat{S}=\exp\Big(-i\sum_{\phi,\phi'}{h}_{I,\phi,\phi'}\ket{\phi}\bra{\phi'} \Big),  \label{eq:S} \\    
     &{h}_{I,\phi,\phi'}=\delta_{E_{\phi},E_{\phi'}}  \frac{L}{\hbar v_0}\bra{\phi}\hat{H}_{I}\ket{\phi'}. \label{eq:Hphi}
\end{flalign}

The Kronecker delta in Equation~\eqref{eq:Hphi} accounts for energy conservation in the target-electron interaction. It is obtained by taking the discrete limit of the continuous delta function~\cite{Zhao2021}, $\delta((E_{\phi}-E_{\phi'})/\hbar) \rightarrow (L/2\pi v_0)\delta_{E_{\phi},E_{\phi'}}$, which is a consequence of the particle-in-a-box quantization of the electron wavepacket. This introduces a discrete resolution in wave-vector $\Delta k=2\pi/L$, and energy $\Delta E_k=h v_0/L$. The $L$ factor in Equation~\eqref{eq:Hphi} cancels with the $1/L$ factors in the expectation values $\bra{\phi}\hat{H}_{I}\ket{\phi'}$ embeded in the coupling strengths in Eqs.~\eqref{eq:Coupling_final_2}-\eqref{eq:Coupling_final_4}, which makes the propagator $\hat{S}(t\rightarrow\infty)$ independent of this auxiliary length scale. By relating the initial and final free-electron momenta through the momentum exchanged with the target, $k =k'-q$, it is possible to write
\begin{equation}
    \delta_{E_{\phi},E_{\phi'}}\approx \delta\big(q v_0\,,\,\omega_c(n_z-n_z')+\omega_{N,\pm}-\omega_{N',\pm'}\big)
\end{equation}
where we have used the notation $\delta_{i,j}=\delta(i,j)$ for clarity.  
Note that in previous works exploring the electron-beam-probing of optical cavities~\cite{DiGiulio2019} and QEs~\cite{Zhao2020}, all the momentum and energy exchanged with the target was in multiples of $\omega_{c,QE}/v_0$ and $\omega_{c,QE}$, since the latter was the only energy scale present in the system. Here, the cavity-QE SC and the resulting polaritonic ladder gives rise to a much more complex landscape of electron-target interactions.\\

\begin{figure}[t]
\includegraphics[angle=0,width=\columnwidth]{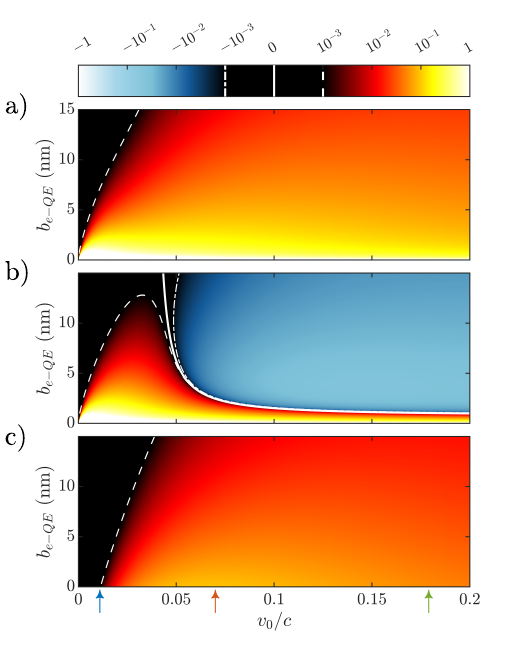}
\caption{Matrix elements, $h_{I,G,\phi'}$, connecting the target ground state with states of the first excitation manifold as a function of the free electron-QE distance and the electron speed.  (a) Upper polariton $\ket{1+}$, (b) Lower polariton $\ket{1-}$, c) $z-$dipolar mode $\ket{1z}$. Note that we are omitting the electronic part of the wavefunction, see main text. Solid, dashed, and dotted-dashed white lines plot the isocurves $h_{I,G,\phi'}=0$, $h_{I,G,\phi'}=10^{-3}$, and $h_{I,G,\phi'}=-10^{-3}$, respectively. The vertical color arrows indicate the configurations considered in~\Cref{fig:fig3}.} \label{fig:fig2}
\end{figure}

\Cref{fig:fig2} shows the adimensional matrix elements ${h}_{I,\phi,\phi'}$ that connect the ground state of the target, $\ket{\phi}=\ket{G}=\ket{0}_z\otimes\ket{0}_\pm\otimes\ket{k}$ and the different states of the first excitation manifold of $\hat{H}_0$. With the cavity and QE parameters introduced above, we obtain $g_x^{c-QE}\approx 80$ meV, which is in accordance with the light-matter interaction strengths reported experimentally in different nanophotonic-based polaritonic  systems at the single QE level~\cite{Santhosh2016,Chikkaraddy2016}. Due to the structure of the interaction Hamiltonian, the evaluation of Equation~\eqref{eq:Hphi} for $\ket{\phi'}=\ket{1\pm}=\ket{0}_z\otimes\ket{1,\pm}\otimes\ket{k-\tfrac{\omega_{1\pm}}{v_0}}$ and $\ket{\phi'}=\ket{1z}=\ket{1}_z\otimes\ket{0}\otimes\ket{k-\tfrac{\omega_{c}}{v_0}}$ yields~\cite{Karnieli2023}
\begin{eqnarray}
{h}_{I,G,1\pm}&=&\frac{L}{\hbar v_0}\left[g^{e-c}_{\omega_{1,\pm}/v_0,x}\pm g^{e-QE}_{x}\right], \label{eq:1pm}\\ 
{h}_{I,G,1z}&=&\frac{L}{\hbar v_0}g^{e-c}_{\omega_c/v_0,z}.\label{eq:1z}
\end{eqnarray}
Equation~\eqref{eq:1pm} illustrates the power of electron beams for the exploration of light-matter SC. In optical-based spectroscopic techniques, which operate under the far-field, laser-like pumping of the polaritonic target, the driving amplitude of the cavity is orders of magnitude larger than the QE. This is a consequence of the dipole mismatch between them, which is $\mu_{c_{x,z}}/\mu_{QE}\simeq40$ for the small nanoparticle in our system (see Section S4 of the SM). In these setups, the polariton population takes place through the cavity, and hence, it is exactly the same (except for dispersion effects) for lower and upper states. When employing a very localized excitation, the electron beam in our case, it is possible to make the absolute value of two terms in Equation~\eqref{eq:1pm} similar through the tuning of the probe parameters that come into play in the interaction with the target. In this regime, one of the polariton states becomes completely dark to the passing electron, enabling the selective probing of the other one, as all the interaction dynamics will occur solely through it.   
 
\Cref{fig:fig2} render $h_{I,G,1+}$ (a), $h_{I,G,1-}$ (b) and $h_{I,G,1z}$ (c), as a function of the electron-QE impact parameter, $b_{e-QE}$, and the central velocity of the electron wavepacket normalized to the speed of light, $v_0/c$. We can observe that all the matrix elements decrease with larger distance and lower velocity (see dashed white lines), although only $h_{I,G,1-}$ completely vanishes within the parameter range considered, as indicated by the white solid line in panel (b). As expected from the setup we have chosen, see \Cref{fig:fig1}, the electron probes more efficiently the polaritonic states than the $z$-dipolar cavity mode at small $b_{e-QE}$. Only at large $v_0/c$, the three panels acquire similar absolute values, although the elements for the lower polariton change sign and become negative in this regime. The study provided in these three panels serves as a guide for the design the electron-beam configuration most appropriate to interrogate a given state of the first excitation manifold in the light-matter SC target.

The adimensional matrix elements in  \Cref{fig:fig2} acquire values that range between -1 and 1, which means that the propagator in Equation~\eqref{eq:S} can be treated perturbatively in different orders of probe-target interaction for most of the configurations analyzed. Using the Taylor expansion for the exponent function, we can write $\hat{S}=\sum_{\phi,\phi'}S_{\phi,\phi'}\ket{\phi}\bra{\phi'}$, with
\begin{flalign}
    S_{\phi,\phi'}=\delta_{\phi\phi'}-ih_{I,\phi,\phi'}-\tfrac{1}{2}\sum_{\phi''}h_{I,\phi,\phi''}h_{I,\phi'',\phi'}+\ldots \label{eq:Staylor} 
\end{flalign} 
which shows explicitly the mixing of the states of $\hat{H}_0$ by the passing electrons to all orders in the coupling strengths given by Eqs.~\eqref{eq:Coupling_final_2}-\eqref{eq:Coupling_final_4}. In the following sections, we will investigate how this electron-induced mixing can be exploited to extract information about the polaritonic states of the target. We will focus first on incoming electrons with a well-defined momentum, and then proceed to explore how modulated electron beams can be used to further characterize light-matter SC phenomena in the target through the engineering of the electron wavefunction.

\section{CL, EELS and PINEM in polaritonic targets}
There are two strategies that allow extracting information from the target through the electron probing: through the radiation spectrum of the cavity (we neglect the emission from the QE) into the far-field , as it is done in CL setups, and through the energy lost/gained by the electron beam itself, like in EELS or PINEM experiments. We consider the former first, whose characterization is given by its power spectrum~\cite{Carmichael89}
\begin{gather}
    I(\omega)=\lim_{T\rightarrow \infty}\frac{1}{2\pi T}\int_{-\frac{T}{2}}^{\frac{T}{2}} dt \int_{-\infty}^{\infty}d\tau \expval{\hat{\xi}^\dagger(t+\tau)\hat{\xi}(t)}e^{-i\omega\tau},\label{eq:Iomega}
\end{gather}
where $\hat{\xi}=\mu_{c_{x,y}}(\hat{a}_x+\hat{a}_z)$ is the dipole moment operator of the cavity (describing the coherent light emission~\cite{Saez-Blazquez2018} by its two degenerate modes) and  $\hat{\xi}(t)=e^{i\hat{H}_0t/\hbar}\hat{\xi}e^{-i\hat{H}_0t/\hbar}$ describes its evolution in time under the bare Hamiltonian in Equation~\eqref{eq:simplifyied_hamiltonian1}. The expectation value in in Equation~\eqref{eq:Iomega} is firstly taken over the state $\ket{\phi_f}=\hat{S}\ket{G}$, which results from the fast target-probe interaction when the former is initially in its ground state. We have briefly discussed the lifetime of the target states to justify the approximations inherent to Equation~\eqref{eq:S}. However, our model is based on a purely Hamiltonian description of the target, given by Equation~\eqref{eq:phi-states}. Therefore, the spectra obtained from Equation~\eqref{eq:Iomega} will consist of a weighted sum of delta Dirac functions. In the following, we will introduce a phenomenological broadening, $\sigma$, for the spectral features, to account for the finite lifetime of the target states, by making the replacement $\delta(\omega)\rightarrow\tfrac{\sigma}{2\pi}\tfrac{1}{\omega^2+\sigma^2/4}$. This lorentzian lineshape is obtained in the Lindbladian description of open quantum systems~\cite{Medina2021,Breuer2007}.

\begin{figure}[!t]
\includegraphics[angle=0,width=\columnwidth ]{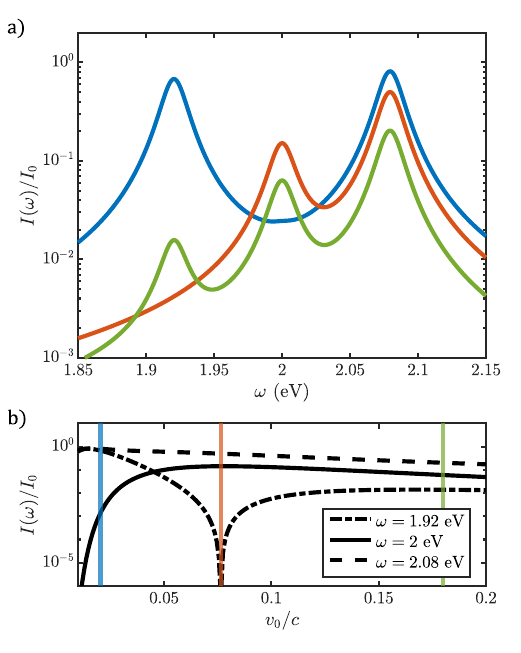}
\caption{Far-field light intensity versus photon frequency for passing electrons with $b_{e-QE}=1$ nm and $b_{e-c}=11$ nm. (a) Power spectra for three different electron velocities, indicated by the vertical arrows in \Cref{fig:fig2}: $0.02c$ (blue), $0.08c$ (orange) and $0.18c$ (green). (b) Height of the three spectral maxima of $I$ spectra as a function of $v_0/c$. Vertical lines indicate the configurations considered in (a).} \label{fig:fig3}
\end{figure}

\Cref{fig:fig3}(a) shows CL-like spectra obtained for aloof electrons with impact parameters $b_{e-QE}=1$~nm, $b_{e-QE}=11$~nm, and different velocities, indicated by the vertical color arrows in \Cref{fig:fig2}. The far-field intensity spectra is broadened by $\sigma$, set to 0.02~eV, an optimistic estimation for plasmonic lifetimes~\cite{Kholmicheva2018} ($1/\sigma=30$~fs). They are normalized to $I_0$, the intensity at the polariton maxima in the limit $v_0\rightarrow0$ (see below). Three spectral maxima are apparent, which originate from the upper and lower polaritons, at 2.08 and 1.92~eV, respectively, and the uncoupled $z-$dipole cavity mode at $2$~eV. For slow electrons (blue, $v_0=0.02c$), the spectrum is dominated by the polariton peaks, which have similar weights. This indicates that the electron-target interaction is mainly taking place through one of the polariton constituents. Indeed, $|g_x^{e-QE}|>>|g_{\omega_{1\pm}/v_0,x}^{e-c}|$ in this case, due to the small value of the QE impact parameter. As expected from \Cref{fig:fig2}(c), there is not an intermediate peak in this spectrum, as $g_{\omega_c/v_0,z}^{e-c}$ is negligible in this configuration.

The spectra for higher electron velocities, $v_0=0.08c$ (orange), does not present the peak at 1.92 eV, which indicates that the lower polariton has become dark to the incoming electron beam. Note that $g_x^{e-QE}\simeq g_{\omega_{1\pm}/v_0,x}^{e-c}$ and $h_{I,G,-}$ vanishes in this case. At even larger velocities, $v_0=0.18c$ (green), the spectral peaks are, in general, lower, but the three of them are clearly visible. In this configuration, all the matrix elements acquire comparable values. Our results reveal the complex dependence of $I(\omega)$ on $v_0/c$, far from any monotonic trend. In \Cref{fig:fig3}(b) we analyze it in more detail, by displaying the height of the CL peaks as a function of the electron velocity. We find that the upper polariton peak is always the largest, while the lower polariton ($z-$dipole mode) is the second largest for low (large) $v_0$. In the limit $v_0\rightarrow0$, the upper and lower polariton maxima acquire the same value, $I_0$, employed for normalization. We can also observe that three far-field intensity maxima approach in the limit of large electron velocity in \Cref{fig:fig3}(b).   

\begin{figure}[!t]
\includegraphics[angle=0,width=\columnwidth ]{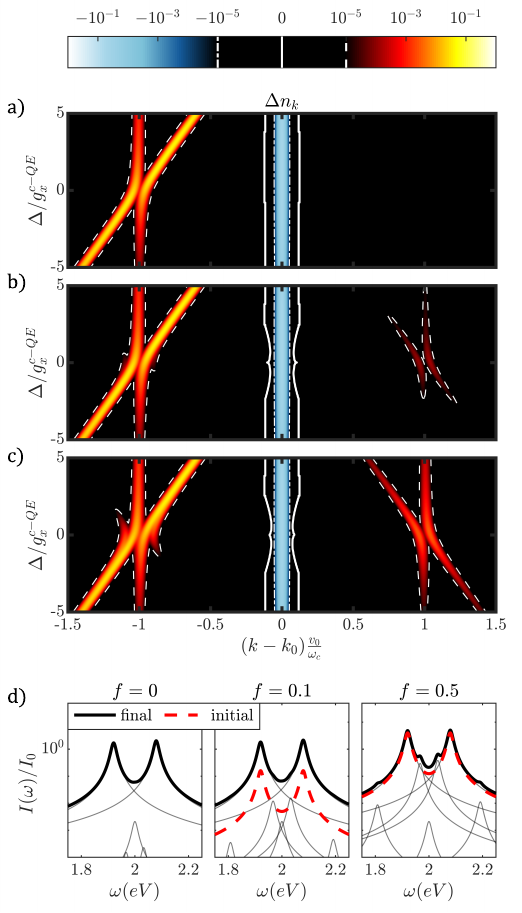}
\caption{Momentum reshaping experienced by an incident monochromatic electron beam ($k=k_0$, $v_0=0.02c$) in its interaction with a polaritonic target as a function of the half cavity-QE detuning $\Delta=(\omega_c-\omega_{QE})/2$. In panel (a), the cavity is initially in its ground state in an EELS-like configuration. In panels (b) and (c), the initial state of the cavity is given by Equation~\eqref{eq:state_EELSPINEM} with $f=0.1$ and 0.5, respectively, mimicking a PINEM setup. Panel (d) shows far-field emission spectra for the targets in the three panels above and $\Delta=0$. Black solid (red dashed) lines plot the intensity after (before) the interaction with the incoming electrons. Thin grey lines render the different contributions to the intensity spectra.} \label{fig:4}
\end{figure}

We investigate next the fingerprint of the target-probe interaction in the wavefunction of the passing electron beam. For this purpose, we focus on the reshaping of the momentum distribution of the electron wavepacket, measured by the difference in the population of the states $\ket{k}$ before and after the coupling with the QE-cavity system. Expressed in terms of the number operator $\hat{n}_k=\hat{c}_k^\dagger\hat{c}_k$, this difference is given by
\begin{equation}
\Delta n_k=\expval{\hat{n}_k}-\expval{\hat{n}_k}^0 \label{eq:nk}
\end{equation}
where the superscript 0 indicates that the expectation value is evaluated for the electron wavefunction prior to the interaction. \Cref{fig:4}(a)-(c) plots this population difference versus $(k-k_0)\tfrac{v_0}{\omega_c}$ (where $k_0$ is the central electron wavevector) and half the detuning between the cavity mode and the QE, $\Delta=(\omega_c-\omega_{QE})/2$ (no longer at resonance in our analysis). The panels correspond to different initial states of the cavity-QE target, parameterized through the variable $f$, the amplitude of the first excited state of the $x$-dipole cavity mode,
\begin{equation}
\ket{\phi_0}=\ket{0}_z\otimes\left[\sqrt{1-f^2}\ket{0}_x +f\ket{1}_x \right]\otimes\ket{g}\otimes\ket{k_0}. \label{eq:state_EELSPINEM}
\end{equation}
This eigenstate of $\hat{H}_0$ for vanishing $g_x^{c-QE}$ mimics the weak, coherent driving of the cavity by a laser field polarized along $x$-direction. Note that, we have used the bare basis above, instead of the polaritonic basis employed in the previous section. 

In \Cref{fig:4}(a), we consider an EELS-like configuration, with the target initially in its ground state, $\ket{\phi_0}=\ket{G}$ ($f=0$). This setup has been previously investigated in the context of polariton formation in nanophotonic systems~\cite{Crai2020,Zouros2020,Karnieli2023}. The incoming electron beam is monochromatic, presenting a single wave-vector component, $k=k_0$ and $v_0=0.02c$ (blue arrow in \Cref{fig:fig2}). We can observe that the electron population is transferred to $k<k_0$, the region of energy loss, while, as expected, the energy gain region ($k>k_0$) remains null. At zero detuning, $\Delta=0$, two maxima in  $\Delta n_k>0$ (yellow color) are apparent, corresponding to the polaritonic states in the first excitation manifold, $\ket{1,\pm}$. These emerge in the region $k-k_0\simeq-\tfrac{\omega_c}{v_0}$. The momentum transfer maxima for non-zero detuning disperse, giving rise to the imprint of the anticrossing profile characteristic of light-matter SC~\cite{Sanvitto2016,Santhosh2016,Chikkaraddy2016,Zhu2023,Karnieli2023} into the electron wavepacket. At $|\Delta|>|g_x^{c-QE}|$, two asymptotic branches are apparent, one vertical, corresponding to the $x$-dipole cavity mode (fixed $\omega_c$), and one diagonal, given by the QE exciton (variying $\omega_{QE}$). Like in \Cref{fig:fig3}, a phenomenological wave-vector broadening $\sigma/v_0$ has been introduced in the map. The resulting lineshapes are indicated by the solid and dashed lines, which correspond to the isocurves $\Delta n_k=0$ and $\Delta n_k=10^{-5}$, respectively. 

As shown in \Cref{fig:4}(b), by pumping weakly the cavity mode ($f=0.1$), and under a monochromatic electron beam with $k=k_0$ and $v_0=0.02c$, a region of $\Delta n_k>0$ emerges in the energy-gain side of the momentum transfer map. This indicates that, as a result of the interaction with the target, the electron wavepacket can acqire momentum components larger than $k_0$ thanks to the population in the first excitation manifold of the cavity. This setup mimics a PINEM experiment, in which the passing electrons exchanges energy with an optically-driven resonator. We can observe that the anticrossing profile in the energy-gain region is the fainted mirror image of the energy-loss one, with asymptotic branches given by fixed $\omega_c$ and $-\Delta$. At higher driving, $f=0.5$ in \Cref{fig:4}(c), the magnitude of the energy gain anti-crossing becomes comparable to its energy loss counterpart, as the amplitude of $\ket{0}_x$ and $\ket{1}_x$ in Equation~\eqref{eq:state_EELSPINEM} are the same. We can also observe extra branches in the energy loss region, that follow $-\Delta$ instead of $\Delta$. These $\Delta n_k$ maxima originate from the promotion of polaritonic population from the first to the second excitation manifold in the interaction with the passing electrons (discussed in more detail below), and illustrates that the power of PINEM in polaritonic systems for electron wavepacket shaping is well beyond that of EELS. 

To complement our study, we plot in \Cref{fig:4}(d) the emission spectrum calculated from Equation~\eqref{eq:Iomega} under the driving conditions in panels (a)-(c) and for zero cavity-QE detuning ($\Delta=0$). Red dashed and black solid lines render $I(\omega)$ (in log scale) before and after the interaction with the electron beam. At $f=0$ (EELS-CL configuration), $I(\omega)=0$ prior to the electron arrival, and the final spectrum is dominated by two maxima originated from the radiative decay of the $\ket{1,\pm}$ polaritons to the ground state. The lineshapes for these two contributions are rendered in thin grey lines. Due to their lower weight, other spectral contributions also plotted in grey thin lines, are not apparent in $I(\omega)$.  The central one corresponds to the $z-$dipole cavity mode (weakly excited by the passing electrons), and the small ones next to it result from the $\ket{2,\pm}$ to $\ket{1,\pm}$ transitions, with frequencies $\omega_{2,\pm}-\omega_{1,\pm}=\omega_{c,QE}\pm(\sqrt{2}-1)g^{c-QE}_x$. 

\Cref{fig:4}(d) also presents intensity spectra for the two optically-driven cavities in panels (b) and (c), evaluated at $f=0.1$ and 0.5, respectively. In both cases, the initial spectra present the two main polaritonic peaks only, whose height increases with $f$. In the final $I(\omega)$, multiple contributions can be identified. Apart from the two main ones, whose amplitude barely varies with respect to $f=0$, and the central $z-$dipole feature which is independent of $f$, we can observe that the weight of the  $\ket{2,\pm}$ to $\ket{1,\pm}$ transitions grow considerably with increasing optical driving. Moreover, two additional side peaks are apparent, due to another set of second-to-first manifold transitions, $\ket{2,\pm}$ to $\ket{1,\mp}$, with frequencies $\omega_{2,\pm}-\omega_{1,\mp}=\omega_{c,QE}\pm(\sqrt{2}+1)g^{c-QE}_x$. These transitions are also behind the extra branches in the energy-loss side branches of \Cref{fig:4}(c) at small detuning. Our results evidence that the alteration of $I(\omega)$ due to the passing electron is negligible for cavities under significant optical pumping, a direct consequence of the weak character of the probe-target interaction~\cite{Zhao2021,Ruimy2021}. Thus, to fully exploit the probing abilities of electron wavepackets, the strength of their coupling to the polaritonic target must be enhanced. In the next section we explore the use of different degrees of freedom of the incoming electron wavefunction for this purpose.                   

\section{Modulated electron beams}
In the previous section, we have shown that, in a PINEM setup, electron wavefunctions with a rich momentum distribution can be generated from monochromatic electron beams through their interaction with a pumped cavity-QE target. By letting the electrons drift after the interaction, the various momentum components separate in space, giving rise to a series of peaked electron wavepackets. These are usually termed as modulated electron beams. Indeed, in recent years, much research attention have focused on different approaches to generate modulated electron wavefunctions through then interaction with optical systems~\cite{GarciadeAbajo2021b,Henke2021,Madan2022}. Here, we explore the probing capabilities that these modulated electrons bring when interacting with a polaritonic system. 

In this section, we will use the same formalism as in the previous one, but for convinience, we will explicitly deal with target and probe degrees of freedom separately. As a starting point, the bare Hamiltonian eigenstates in Equation~\eqref{eq:phi-states} as $\ket{\phi}=\ket{\varphi_e}\otimes\ket{\psi_t}$, where the first (second) wavefunction characterizes the electron beam (target) state. Similarly, the interaction Hamiltonian in Equation~\eqref{eq:simplifyied_hamiltonian2} can be written as $\hat{H}_I=\sum_q\hat{H}_{I,q}\,\hat{b}_q$, separating target and electron operators, and the scattering matrix in Equation~\eqref{eq:S} as
\begin{eqnarray}
     \hat{S}&=&\exp\Big(-i\sum_{\psi_i,\psi_j} \mathcal{h}_{I,\psi_i,\psi_j} \dyad{\psi_i}{\psi_j}\,\hat{b}_{q_{\psi_i,\psi_j}} \Big)=  \nonumber \\
     &=&\sum_{\psi_i,\psi_j} \mathcal{S}_{\psi_i,\psi_j} \dyad{\psi_i}{\psi_j}\,\hat{b}_{q_{\psi_i,\psi_j}}, \label{eq:Spsi} 
\end{eqnarray}
with $\mathcal{h}_{I,\psi_i,\psi_j}=\tfrac{L}{\hbar v_0}\bra{\psi_i}\hat{H}_{I,q_{\psi_i,\psi_j}}\ket{\psi_j}$, and where $q_{\psi_i,\psi_j}=(E_{\psi_i}-E_{\psi_j})/ \hbar v_0$ is the  target-electron momentum exchange (set by energy conservation and the non-recoil approximation). The scattering matrix amplitudes $\mathcal{S}_{\psi_i,\psi_j}$ have the same form as Equation~\eqref{eq:Staylor}, but replacing the states $\ket{\phi}$ by $\ket{\psi}$, and the matrix elements $h_{I}$ by the $\mathcal{h}_{I}$ ones above. The structure of the scattering matrix in Equation~\eqref{eq:Spsi} allows its analytical implementation through the exploitation of the simple algebra of the $\hat{b}_q$ operators, as detailed in section S6 of the SM. This is where the power of our approach resides, as it makes it possible to obtain analytical expressions for the observables of interest.
\begin{figure*}[t!]
\makebox[\textwidth][r]{\includegraphics[width=1.02\textwidth]{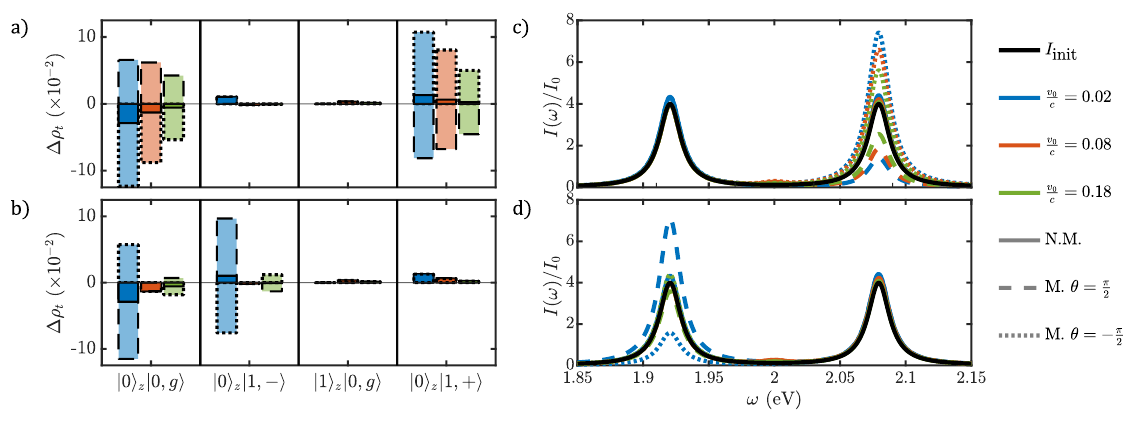}}
\caption{Impact of the electron modulation on the target population transfer (a)-(b) and cavity power spectrum (c)-(d). In the top (bottom) panels, the momentum modulation is at resonance with the transition between the ground state and upper (lower) polariton state, $q_{mod}=\omega_{1,+}/v_0$ ($q_{mod}=\omega_{1,-}/v_0$). Three different electron central velocities are considered: $0.02c$ (blue), $0.08c$ (orange) and $0.2c$ (green), and the impact parameter $b_{e-c}$ is set to 11 nm. In all panels, two different initial state phases, $\theta$ are considered: $\pi/2$ (dashed lines) and $-\pi/2$ (dotted lines). The solid lines correspond to a non-modulated (N.M.) electron beam, and the solid black lines in (c) and (d) plot the cavity spectrum before the interaction with the passing electrons.} \label{fig:fig5}
\end{figure*}

We consider now a modulated electron beam, initially prepared in a superposition of momenta of the form ${\ket{\varphi_e}=\sum_k B(k)\ket{k}}$ (with $\sum_k |B(k)|^2=1$). This wavefunction can describe, for instance, a comb with a set of amplitude peaks equally spaced in momentum space. We assume that the target is prepared initially in one of its polaritonic eigenstates, $\ket{\psi_m}=\ket{n}_z\otimes\ket{N,\pm}$. If, for convenience, we employ its density matrix description of the target, we have $\rho_t^{0}=\ket{\psi_m}\bra{\psi_m}$. After the interaction with the modulated electrons, it reads
\begin{flalign}
     \rho_{t}&=\sum_k \bra{k}\hat{S}\ket{\varphi_e}\rho_t^0\bra{\varphi_e}\hat{S}^\dagger\ket{k}= \nonumber \\
    &=\sum_{\psi_i,\psi_j} \mathcal{S}_{\psi_i,\psi_m} \mathcal{S}_{\psi_j,\psi_m}^* \dyad{\psi_i}{\psi_j}\Big[  \sum_k B(k)B^*(k-q_{\psi_i,\psi_j})\Big].
\end{flalign}
The expression above shows that the population of the target states are completely independent from the electron momentum distribution~\cite{Kfir2021}, as $\bra{\psi_s}\rho_{t}\ket{\psi_s}=|\mathcal{S}_{\psi_s,\psi_m}|^2$. Furthermore, $\bra{\psi_s}\rho_{t}\ket{\psi_s}\simeq\bra{\psi_s}\rho_{t}^0\ket{\psi_s}=\delta_{s,m}$ ($\mathcal{S}_{\psi,\psi}=1$ to first order in the electron-target interaction), which shows that initial populations remain largely unchanged after the interaction with the electron. On the contrary, the coherences can be  manipulated by appropriately designing the electron wavefunction~\cite{Kfir2021}. Thus, by shaping $B(k)$ as a finite momentum comb with spacing $q_{\psi_{m_1},\psi_{m_2}}$, the coherence $\bra{\psi_{m_1}}\rho_{t}\ket{\psi_{m_2}}$ will be modified, while leaving the rest of the target density matrix unaltered.

Next, we focus our attention on targets prepared in a superposition of polaritonic states of the form $\ket{\psi_t^0}=\cos{\phi}\ket{\psi_{m_1}}+e^{i\theta}\sin{\phi}\ket{\psi_{m_2}}$. Then, the population of a given polaritonic state $\ket{\psi_s}$ after interaction with the modulated electron beam has the form 
\begin{eqnarray}
    \bra{\psi_s}\rho_{t}\ket{\psi_s}&=& \cos^2{\phi}\left|\mathcal{S}_{\psi_s,\psi_{m_1}}\right|^2 +\sin^2{\phi} \left|\mathcal{S}_{\psi_s,\psi_{m_2}} \right|^2\nonumber\\
    &&+ \mbox{Re}\Big\{e^{-i\theta}\sin{2\phi}\,\mathcal{S}_{\psi_s,\psi_{m_1}}\mathcal{S}^*_{\psi_s,\psi_{m_2}}\times \nonumber \\ 
    && \sum_k B(k) B^*(k-q_{\psi_{m_2},\psi_{m_1}}) \Big\}, \label{eq:Final_pop_mod_electrons}
\end{eqnarray}
which explicitly shows that for arbitrary initial target state, the final polaritonic populations can vary thanks to the coherences in $\rho_t^0$ (before the target-probe interaction) and the modulation of the electron beam encoded in $B(k)$. See more details in Section S5 of the SM. The last term indicates that by targeting the transition between the polaritonic states involved in $\ket{\psi_t^0}$, the impact of the modulation on the populations can be maximized. This ability (showcased here for a general target) of modulated electrons to transform coherences into populations is what enables them to induce Rabi dynamics in  QEs~\cite{Zhao2020} and makes it possible using them to implement quantum state tomography protocols \cite{Ruimy2021}.     

To illustrate the implications of Equation~\eqref{eq:Final_pop_mod_electrons}, we consider a particular target-probe configuration. The initial electron wavefunction is set to a comb of the form $\ket{\psi_e}=\sum_{n=-N/2}^{N/2}\frac{\ket{k_0+n q_{mod}}}{\sqrt{N+1}}$ with $N=100$. Note that it implies the exchange of up to 50 photons in its preparation (well within reach of recent PINEM experiments \cite{Kaminer2020}). The polaritonic target is initially in the state 
\begin{equation}
\ket{\psi_t^0}=\frac{1}{2}\ket{0}_z\otimes\big[\sqrt{3}\ket{0}_x+e^{i\theta} \ket{1}_x\big]\otimes\ket{g}, \label{eq:inistate}
\end{equation}
where $\theta$ is a real number. Note that this wavefunction can be expressed as a linear combination of the target ground state and the $\ket{0}_z\otimes\ket{1,\pm}$ polaritonic states. 

In \Cref{fig:fig5}(a-b), we analyze the population differences (given by the diagonal terms of $\Delta\rho_t=\rho_t-\rho_t^0$) induced by the passing electrons on the ground and the three first-excitation target states. We consider the three central electron velocities indicated in \Cref{fig:fig2} (blue, orange and green in increasing $v_0/c$), and two different initial state configurations, given different values of $\theta$ in Equation~\eqref{eq:inistate} (dashed and dotted lines). In all cases, $b_{e-c}=11$ nm. For reference, the population differences for a non-modulated (N.M.) electron beam are plotted in solid lines (note that these are independent of $\theta$). In panel (a), the modulation spacing is at resonance with the upper polariton, $q_{mod}=\omega_{1,+}/v_0$, in panel (b), with the lower one, $q_{mod}=\omega_{1,-}/v_0$. 

As expected, \Cref{fig:fig5}(a) displays a significant population transfer only between the ground state (left) and the upper polariton (right), which is larger for lower electron velocity, following the monotonic dependence in the emitter-target coupling in \Cref{fig:fig2}(a). Moreover, we can observe that for $\theta=-\pi/2$ the upper polariton gains population (as in the non-modulated case), while it gets depopulated for $\theta=\pi/2$. Note that this parameter sets the phase, and therefore the sign, of the contribution of the initial coherences to the final populations given by the last term in Equation~\eqref{eq:Final_pop_mod_electrons}. We can see how this can be leveraged to control the flow of population among polaritonic states. The momentum spacing in $B(k)$ is set to yield the most efficient energy transfer between the ground and lower polariton states in \Cref{fig:fig5}(b). The non-monotonic dependence of the populations on the electron velocity in this case is inherited from \Cref{fig:fig2}(b). Again, varying $\theta$ inverts the direction of the population transfer.    

Apart from analyzing the effect of electron modulation on the target populations, we also investigate its impact on the cavity power spectrum given by Equation~\eqref{eq:Iomega}, now evaluated for the state that results from applying the scattering matrix on Equation~\eqref{eq:inistate}. Importantly, this is a far-field magnitude that can be easily accessed experimentally. \Cref{fig:fig5}(c) and (d) plot $I(\omega)$ for $q_{mod}=\omega_{1,+}/v_0$, and $q_{mod}=\omega_{1,-}/v_0$, respectively. The black solid line renders the cavity spectrum before the interaction with the electron beam, $I_{\rm init}$. We can observe that only the upper polariton peak is shaped by the passing electrons in (c), and the lower polariton one in (d). This illustrates the far-field fingerprint of the population manipulation in panels (a) and (b). In both cases, only the emission from the targeted transition through $q_{mod}$ is modified, keeping the spectrum around the other features unaltered. Importantly, as we observed in the polariton populations, the initial coherences, whose contribution to the spectrum depends on $\theta$, set whether the altered emission peak increases or decreases with respect to $I_{\rm init}$.  

\Cref{fig:fig5} indicates that the coherences, rather than the populations, in $\rho_t^0$ dictate the manner in which the population transfer and the spectrum reshaping take place through the interaction with the modulated electron beam. To gain insight into this result, we simply evaluate Equation~\eqref{eq:Final_pop_mod_electrons} for polaritonic states that are initially populated, i.e., by making $s=m_1$, for example. It is then straightforward to see then that, in the first two terms, $|\mathcal{S}_{\psi_{m1},\psi_{m1}}|^2=1$ and $|\mathcal{S}_{\psi_{m1},\psi_{m2}}|^2=|\mathcal{h}_{I,\psi_{m1},\psi_{m2}}|^2$ to first order in the electron-target interaction. On the contrary, we have $\mathcal{S}_{\psi_{m1},\psi_{m1}}\mathcal{S}^*_{\psi_{m1},\psi_{m2}}=\mathcal{h}_{I,\psi_{m1},\psi_{m2}}$ to first order in the last one. Thus, we find that, while the first terms are independent or quadratic on the electron-target interaction strength, the last is linear, which makes it the leading one. Moreover, it is easy to show that $\sum_k B(k) B^*(k-q_{mod})=N/(N+1)\simeq 1$ for a finite, but long, electron comb, also contributing to make the initial coherences crucial in establishing the effect of the passing electrons on the target. Note that adding a running phase difference between the different momentum peaks of the initial electronic wavefunction as $\ket{\psi_e}=\sum_{n=-N/2}^{N/2}e^{i n \xi}\frac{\ket{k_0+n q_{mod}}}{\sqrt{N+1}}$ yields an extra phase factor in the modulated contributions of Equation~\eqref{eq:Final_pop_mod_electrons}, which allows to externally control the internal dynamics of the target. 

\begin{figure}[h]
\includegraphics[angle=0,width=\columnwidth ]{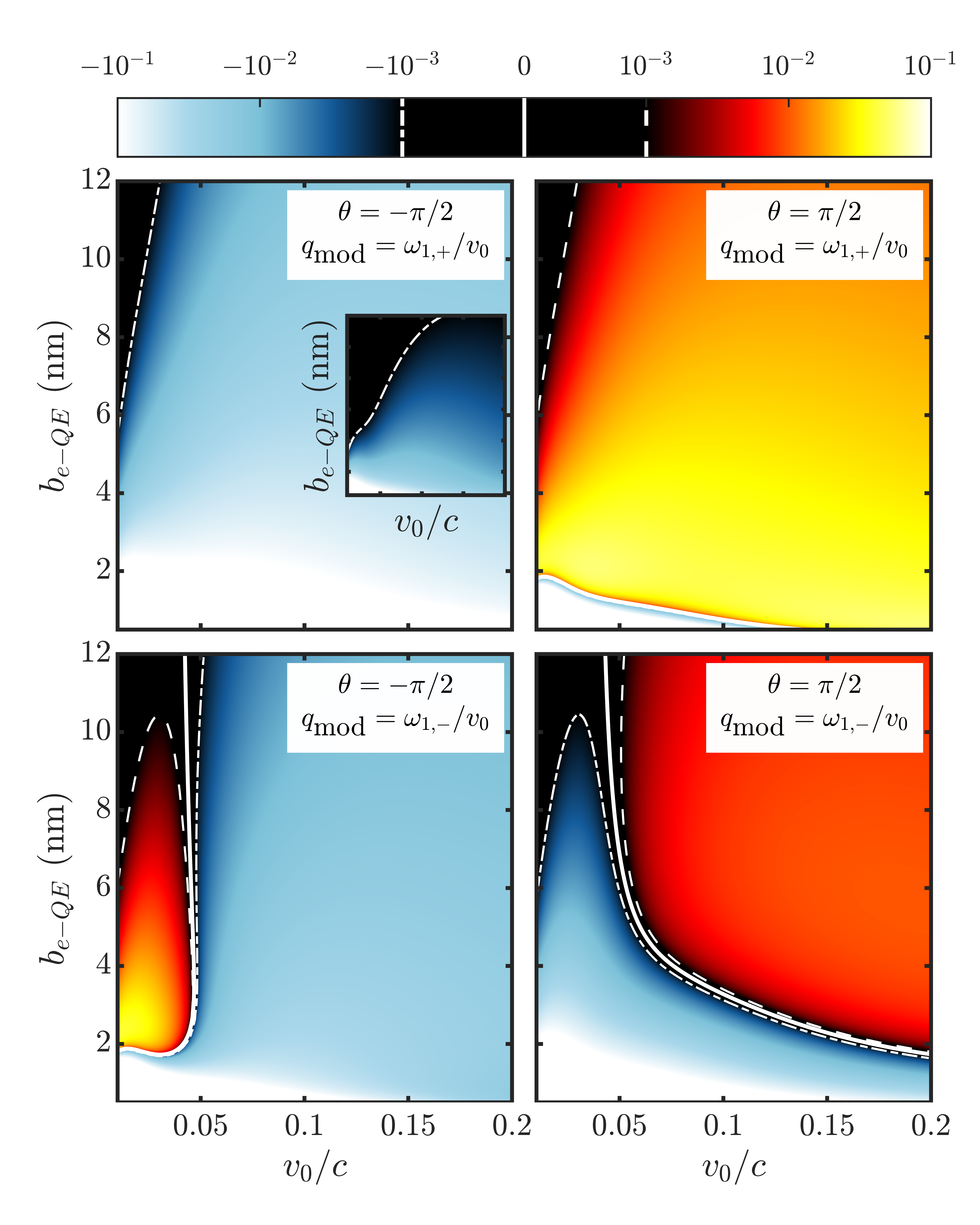}
\caption{Expectation value of the energy change of the electron in units $\hbar \omega_c$. Main panels show the case of modulated electrons, showcasing that net energy gain and loss is achievable by modulating the electron. On each panel we show the modulation spacing and also the phase factor of the initial target state. The inset corresponds to the case of a non-modulated electron, where there always is net energy loss. The result in this case is independent of the phase factor. } \label{fig:fig6}
\end{figure}

Finally, we pay attention to the effect that the probe-target interaction has on the modulated electron beam. The fact that the population transferences induced by modulated beams are larger than the non-modulated ones means that the energy balance of the interaction can be altered through the modulation itself. Thus, it is possible, in principle, to pump or deplete the target. In \Cref{fig:fig6}, we explore the net energy change experienced by the passing electrons 
\begin{equation}
\Delta E =  \sum_k E_k \Delta n_k,  
\end{equation}
where $\Delta n_k$ is defined in Equation~\eqref{eq:nk}. As the initial electron wavefunction, we take the finite comb in  \Cref{fig:fig5} and the target is prepared in the state given by Equation~\eqref{eq:inistate}. 

The four panels in \Cref{fig:fig6} display $\Delta E$ in units of $\hbar\omega_c$ as a function of the central electron velocity and impact parameter, $b_{e-QE}$. The results for the initial target state with $\theta=-\pi/2$ ($\theta=\pi/2$) are shown in the left (right) maps, and the modulation is set at resonance with the ground transition to the upper (top) and lower (bottom) polariton. For reference, the map for non-modulated electrons is shown as an inset with the same parameter range, illustrating that the passing electrons can only lose energy in the non-modulated setup, and $\Delta E$ is larger for smaller impact parameter and electron velocity. The situation is rather similar for $\theta=-\pi/2$. For this state phase, there emerges only a narrow region of small $v_0/c$ where the electron beam gains energy for $q_{mod}=\omega_{1,-}/v_0$. Apart from it, the maps resemble the EELS one, and the target populations always increases by the effect of the passing electrons. For $\theta=\pi/2$ (right), the net energy change maps are very different. Fast electron beams gain energy for both $q_{mod}$ (although $\Delta E$ is larger for the transition between the ground and upper polariton), and lose it at low velocities and impact parameters. Here, the target is populated/depopulated depending on $v_0$ and $b_{e-QE}$. The richness of the net energy loss/gain landscape in \Cref{fig:fig6} follows from the coupling strengths in \Cref{fig:fig2}, as the leading order in the electron-target interaction is linear in $h_{I}$. Thus, we can link the gain-loss transitions in the lower maps with the change in sign in $h_{I,G,1-}$ in \Cref{fig:fig2}(b). All maps are equivalent in the limit of small $v_0$ and $b_{e-QE}$, as $h_{I}\rightarrow 1$ in this limit and the electron modulation becomes irrelevant. Our results also showcase the power of polaritonic systems to re-shape and alter modulated electron beams through the energy of its natural transitions and the phase involved in its initial state preparation.          

\section{Conclusions}
We have presented a comprehensive study of the probing of polaritonic systems by electron beams. The target is composed by a nanophotonic cavity supporting two dipolar modes, and a quantum emitter strongly coupled to one of them. Using macroscopic QED, we have built a model Hamiltonian describing the interaction between probe and target, fully parameterized in terms of the Dyadic Green's function in the quasi-static approximation. We have analyzed the effect of electron-polariton interactions on different observables, including the electron momentum distribution and net energy change, and the polaritonic state populations and the light emission spectrum by the target. Our investigation has proceeded by increasing the complexity on the electron beam and target preparation, from EELS and CL to PINEM, and finally PINEM with modulated electron beams. All these described using the same, unifying theoretical model. Our results show that free electrons, through the modulation of their wavefunction, are a powerful probe, and also a suitable tool for the manipulation, of quantum targets with a complex energy ladder of (bright and dark) excitations, such as polaritonic systems. \\


\begin{acknowledgements}
The authors thank Francisco J. Garc\'ia-Vidal for fruitful discussions and acknowledge funding from the Spanish Ministry of Science, Innovation and Universities through Grants Nos. PID2021-126964OB-I00 and TED2021-130552B-C21, as well as the European Union’s Horizon Programme through grant 101070700 (MIRAQLS) and the Proyecto Sin\'ergico CAM 2020 Y2020/TCS-6545 (NanoQuCo-CM).
\end{acknowledgements}

\bibliography{bib_electrons}


\end{document}



\title{Supplementary Material:\\
Electron-assisted probing of polaritonic light-matter states}

\author{J. Abad-Arredondo}
\email{jaime.abad(at)uam.es}
\author{A. I. Fern\'andez-Dom\'inguez}%

\affiliation{Departamento de F\'isica Te\'orica de la Materia
Condensada and Condensed Matter Physics Center (IFIMAC),
Universidad Aut\'onoma de Madrid, E- 28049 Madrid, Spain 
}%


\begin{abstract}
This Supplementary Material provides details on different aspects of the theoretical approach presented in the main text. Section S1 showcases a general overview of the construction of the Hamiltonian. Sections S2 to S4 describe  the individual steps taken in its development. Section S2 extends previous frameworks for the study the field-assisted interaction between generic electronic transitions, starting from the Macroscopic QED formalism. Section S3 applies the ideas of emitter-centered modes to define physically meaningful bosonic operators for these general electronic transitions. In Section S4, we particularize the general couplings obtained in previous sections to the interaction with a metallic nanoparticle, by deriving its Dyadic Green's function in the quasi-static limit. A general calculation of the final populations of any quantum target after the interation with a modulate electron are given in Section S5, and Section S6 provides insight into the symbolic computational implementation of our calculations.
\end{abstract}

\keywords{Quantum emitter, Cavity mode, Polaritonic State, Modulated electron beam, Spectroscopy, Microscopy}


\maketitle
 \newpage 

\tableofcontents

\clearpage

\section{Outline of the hamiltonian derivation} \label{SM:Hamiltonian_derivation_outline}
In this section, we give an overview of the way the hamiltonian of the system has been postulated. For a more detailed description of the followed procedure we refer the reader to the corresponding sections of this Supplementary Material document. The starting point is a Macroscopic QED (MQED) hamiltonian describing EM fields in a minimal coupling scheme and a term that describes a collection of arbitrary electronic eigenstates:
\begin{eqnarray}
    \hat{H}=\sum_\lambda \int \int dr\, d\omega\,   \hbar \omega \hat {f}_\lambda^\dagger(r,\omega)\hat {f}_\lambda(r,\omega) +\sum_i E_i \hat{c}_i^{\dagger}\hat{c}_i  + \frac{e}{m} \hat{p}\cdot \hat{A},
\end{eqnarray}
where ${f}_\lambda(r,\omega)$ are the usual bosonic operators representing the fields created by an infinitesimal dipole moment, $\hat{c}_i$ is the annihilation operator of an electron in an eigenstate described by the wavefunction $\phi_i(r)$, and $\hat{p}$ and $\hat{A}$ are the usual momentum and vector potential operators. In the above hamiltonian, all the interactions between different electronic transitions are mediated by EM fields, and as such these interactions are a second order process. The first step is to derive an effective hamiltonian that explicitly couples different electronic transitions. Following a procedure similar to \cite{Dung2002} (described in Section \ref{appendix:matter-matter_interaction_hamiltonian_a_la_Dung}) we find that the dynamics of the system can be described by an effective hamiltonian
\begin{flalign}
\hat{H}_{\mbox{eff}}= \sum_\lambda \int \int dr\, d\omega\,   \hbar \omega \hat {f}_\lambda^\dagger(r,\omega)\hat {f}_\lambda(r,\omega)
    +& \sum_i \left(E_i - \hbar \delta_i \right)\hat{c}_i^\dagger \hat{c}_i \nonumber\\
    +&\frac{e}{m} \hat{p}\cdot \hat{A}'  -\hbar \sum_{\substack{i,j\\ k,l}}g_{ij,kl}^{m-m}\sigma_{ij}\sigma_{kl}^{\dagger}
\end{flalign}
where $\sigma_{ij}=\hat{c}_i^\dagger \hat{c}_j$ is an operator describing the electronic transition $\phi_j\rightarrow\phi_i$, and the expressions for the Lamb-shift, $\delta_i$ and coupling between electronic transition $g_{ij,kl}^{m-m}$ are given by
\begin{eqnarray}
&g_{ij,kl}^{m-m}=\frac{e^2 \hbar \mu_0}{2 m^2} 
\times  \iint  d^3 r\,  d^3 & r'\,   \vec{d}_{ij}(r) \cdot \mbox{Re}\left\{\overline{\overline{G}}(r,r',\Omega)\right\}\cdot \vec{d}_{kl}^*(r'), \label{eq:coupling_matter-matter_0}\\
&\delta_i=\sum_j g_{ij,ij}^{m-m},
\end{eqnarray}
where $\vec{d}_{ij}(r)=\phi_i^*(r)\nabla\phi_j(r)$ is a dipole moment density that comes from expanding the momentum operator in the basis of electronic eigenstates and $\overline{\overline{G}}(r,r',\Omega)$ is the classical dyadic Green's function of the EM problem. One of the assumptions needed for the above expression to hold is that the energy difference associated with transitions $i\rightarrow j$ and $k\rightarrow l$ are similar so that $\Omega = \omega_{ij}\approx\omega_{kl}$. This effective hamiltonian would also hold in the case that the Greens function varies sufficiently smoothly over the frequency range of interest. Note that since $\vec{d}_{ij}^*=-\vec{d}_{ji}$, and the dyadic is symmetric with respect to index exchange, it can be shown that this coupling satisfies: $g_{ij,kl}^{m-m}=(g_{kl,ij}^{m-m})^*$ and $g_{ij,kl}^{m-m}=g_{lk,ji}^{m-m}$.\\

Once that we have included a direct field-mediated coupling between electronic transitions, we proceed to introduce new bosonic operators describing the EM field in the spirit of the emitter-centered modes described in \cite{Feist2020}. This is done in Section \ref{appendix:light-matter_interaction_hamiltonian_a_la_Johannes}, and the final hamiltonian reads
\begin{flalign}  
\hat{H}_{\mbox{eff}}= \sum_{ij} \int d\omega\,   \hbar \omega\, \hat {a}_{ij}^\dagger(\omega)\hat {a}_{ij}(\omega)
    + \sum_i \left(E_i - \hbar \delta_i \right)\hat{c}_i^\dagger \hat{c}_i \nonumber \\
    +\hbar \sum_{ij}\int d\omega \, g_{ij}^{l-m}(\omega)\left[\hat{a}_{ij}(\omega)+\hat{a}_{ji}(\omega)^\dagger\right]\sigma_{ij} \nonumber \\
    -\hbar \sum_{\substack{i,j\\ k,l}}g_{ij,kl}^{m-m}\sigma_{ij}\sigma_{kl}^{\dagger},\label{eq:eff_hamiltonian_general}
\end{flalign}
where the new set of bosonic operators $\hat{a}_{ij}$ is defined as
\begin{flalign}
    \hat{a}_{ij}(\omega)=\frac{-e}{m\omega g_{ij}^{l-m}(\omega)} \iint  d \mathbf{r} \,  d \mathbf{r}'\, \vec{d}_{ij}(\mathbf{r}) \cdot   \left( \sum_\lambda   \overline{\overline{G}}_\lambda(\mathbf{r},\mathbf{r}',\omega )  \cdot  \hat{f}_\lambda(\mathbf{r}',\omega)\right),
\end{flalign}
and satisfy  the canonical conmutation relation of bosonic operators by construction: $[\hat{a}_{ij}(\omega),\hat{a}_{ij}^\dagger(\omega)]=1$, which leads to the following expression for the couplings 
\begin{flalign}
    g_{ij}^{l-m}(\omega)&=\frac{e}{m} \sqrt{\frac{\hbar \mu_0}{\pi}}\sqrt{\iint dr\, dr'\,  \vec{d}_{ij}(r)\cdot \mbox{Im}\Big\{\overline{\overline{G}}(r,r',\omega)\Big\} \cdot \vec{d}_{ij}^*(r')}.\label{eq:coupling_light-matter_0}
\end{flalign}
Since the bosonic operators are constructed to comply with canonical conmutation relations, the couplings that derive from these are always physical, and independent of the particular system under treatment. We remark that until this point, no knowledge about the particular electronic transitions under study have been assumed, and therefore this treatment can be applied to a wide variety of systems beyond the present one. In Section \ref{SM2:electronic_transitions_of_interest}, we find that the current densities,$\vec{d_{ij}}(r)$, associated to transitions in dipolar QE and free electrons propagating along the $\hat{z}$ axis are given by:
\begin{gather}
    \vec{d}_{ij}^{QE}(\mathbf{r})=-\frac{m\omega_{ij}}{e\hbar}\vec{\mu}_{ij}\delta^3(\mathbf{r}-\mathbf{r}_{0}), \\
    \vec{d}_{ij}^{e}(\mathbf{r})=i k_0\frac{\delta^2(\mathbf{r}-\mathbf{r}_{\perp})}{L}e^{i(k_j-k_i)z} \hat{z},
\end{gather}
where $\vec{\mu}_{ij}=-e\bra{j}\hat{r}\ket{i}$ is the usual transition dipole moment,  $k_0 = m v_0 /\hbar$ is the initial momentum of the incoming electron and $L$ is the length of a fictitious box used to quantise the momentum values of the electron. We have also introduced $\mathbf{r}_{0}$ as the location of the dipolar QE and $\mathbf{r}_{\perp,0}$ for the location of the free electron on the $x-y$ plane. To obtain the light-matter interaction coupling strengths, we derive the Green's function for a metallic sphere in Section \ref{sec_sm:QS_Sphere_GF_derivation}, and noting that it can be written as $\overline{\overline{G}}(\mathbf{r},\mathbf{r}',\omega)\approx \overline{\overline{\mathcal{G}}}(\mathbf{r},\mathbf{r}',\omega_C)\delta(\omega-\omega_C)$, where $\omega_C$ is the resonance frequency of the cavity, the coupling strength and interaction hamiltonian for dipolar QE interacting with a single cavity mode are given by:
\begin{flalign} 
    \hat{H}_{I}^{c-QE}=&\hbar g^{c-QE} \left[  \hat{\sigma} \hat{a}^\dagger + \hat{\sigma}^\dagger\hat{a}\right], \\
    g^{c-QE}&=\frac{\left\lvert \omega_{QE}\right\rvert}{\hbar}\sqrt{\frac{ \hbar\mu_0}{\pi}\boldsymbol{\mu}\cdot \Im\Big\{\overline{\overline{\mathcal{G}}}(\mathbf{r}_0,\mathbf{r}_0,\omega_c)\Big\} \cdot \boldsymbol{\mu}^*}.
\end{flalign} 
The spherical nanoparticle supports three degenerate dipolar modes that can be considered to be oriented along the $\hat{x}$, $\hat{y}$ and $\hat{z}$ axis respectively. The coupling between a QE, placed along the $\hat{x}$ axis to each of these degenerate modes is given in Equations \ref{eq:MQED_dip_dip_cavity_coupl_x}-\ref{eq:MQED_dip_dip_cavity_coupl_z}. In the case we are considering, with the dipole moment of the QE's transition oriented along the $\hat{x}$ axis, then only the coupling to the $\hat{x}$-mode of the cavity is non-zero and is given by:
\begin{gather*}
    g^{c-QE}_x=\frac{\left\lvert \omega_{QE}\right\rvert}{3}\sqrt{ \frac{\pi}{2}\left( \frac{R}{b_{c-QE}}\right)^3 \frac{ 1}{\hbar\omega_{c}} \,\frac{\mu_{QE}^2}{\epsilon_0\,b_{c-QE}^3}}.
\end{gather*}
On the other hand, the hamiltonian describing the coupling between free electrons and the cavity is given by
 \begin{flalign}
    \hat{H}_I^{e-c}&=\hbar\sum_{q } g_{q}^{e-c}  \hat{b}_q \left(\hat{a}^\dagger-\hat{a}\right) \mbox{sign}(q), \\
    g_{q}^{e-c}&=\frac{e k_0}{mL}\sqrt{\frac{\hbar \mu_0}{\pi}\iint dz\, dz'\, \Im\Big\{\hat{z}\cdot\overline{\overline{\mathcal{G}}}(\mathbf{R},\mathbf{R}',\omega_c) \cdot\hat{z}\Big\} e^{iq(z-z')} },
\end{flalign}
where for compactess we have defined $\mathbf{R}=[\mathbf{r}_{\perp,0},z]$ and $\mathbf{R}'=[\mathbf{r}_{\perp,0},z']$. Introducing the GF of the cavity we find that the free-electron-cavity couplings can be written as
\begin{flalign}
    g_{q,x}^{e-c}&=\frac{e\hbar  k_0}{3mL} \left\lvert q\right\lvert^2  K_1(\left\lvert q b_{e-c} \right\lvert) \sqrt{\frac{1}{\epsilon_0 \hbar\omega_{c}}\frac{\pi R^3}{2}  },\\
    g_{q,y}^{e-c}&=0,\\
    g_{q,z}^{e-c}&=\frac{e\hbar k_0}{3mL} \left\lvert q\right\lvert^2 K_0(\left\lvert q b_{e-c}\right\lvert)\sqrt{\frac{1}{\epsilon_0 \hbar \omega_{c}}\frac{\pi R^3}{2}   },
\end{flalign}
which shows that the $\hat{y}$ dipolar mode of the cavity is decoupled from the electron and QE and therefore remains outside of the dynamics, while the $\hat{x}$ mode couples to free-electron and QE and the $\hat{z}$ mode couples only to the free electron. 

Finally, the interaction between the free electron and dipolar transition inside the QE can be described through the matter-matter interaction term in the MQED hamiltonian. Upon introducing the particular definition of the current densities associated to both transitions, the interaction hamiltonian and  coupling strength are given by:
\begin{flalign}  
\hat{H}_I^{e-QE}&=-\hbar\sum_{q}g_{q}^{e-QE}\left[\hat{\sigma}-\hat{\sigma}^\dagger\right] \hat{b}_q,\\
g_{q}^{e-QE}&=i k_0\frac{e \mu_0 \omega_{QE}}{ m L} \int d z'\,\boldsymbol{\mu}\cdot \mbox{Re}\left\{\overline{\overline{G}}\left(\mathbf{r}_{0},\mathbf{R}',|\omega_{QE}|\right)\right\}\cdot \hat{z} e^{i q z'}.
\end{flalign}
Upon realizing that this coupling strength depends on the real part of the GF, and that in resonance the cavity's GF is purely imaginary, then the QE-free electron interaction will be mediated by the vacuum's GF. Upon integration the coupling for an arbitrary dipole orientation is given by Equation \ref{eq:coupling_free_electron-bound_electron}. Particularizing for a QE oriented along the $\hat{x}$ direction yields a coupling strength:
\begin{gather}
g_{q}^{e-QE}= -\frac{e\, k_0 \left\lvert q \right\lvert^2 }{2 \pi m L  \epsilon_0\omega_{QE}}    \mu_{QE} \mbox{sign}(q b_{e-QE}) K_1(\left\lvert q b_{e-QE}\right\lvert)
\end{gather}
Putting all these interaction terms together, and introducing the energies for the bare optical modes $\hbar\omega_c \hat{a}_x^\dagger \hat{a}_x$, $\hbar\omega_c \hat{a}_z^\dagger \hat{a}_z$, electronic eigenstates $\sum_i E_i\hat{c}_i^\dagger \hat{c}_i$ and bare QE energies $\hbar \omega_{QE} \hat{\sigma}^\dagger\hat{\sigma}$ we have the complete hamiltonian from the main text.\\

To end this section, we remark that the interaction between a free electron and optical cavity and QE have been studied before (mostly as separate processes, but recently also in conjunction), and their coupling parametrized. However, in this work we have managed to unify their description in terms of the common starting framework of MQED. This ensures proper normalization of optical modes in arbitrary EM environments, and therefore, physically meaningful coupling stregths, even in situations in which the EM environment only serves as a non-resonant background. This work paves the way for exploring quantum phenomena in field mediated interactions between arbitrary electronic transitions, and in general light matter interaction phenomena beyond the dipolar approximation of quantum emitters.

\section{General light matter interaction hamiltonian for arbitrary electronic transitions}
\subsection{Matter-matter interaction through light} \label{appendix:matter-matter_interaction_hamiltonian_a_la_Dung}

From a minimal coupling scheme within a MQED formalism, one is able to describe the coupling between electronic transitions and field photons quite easily. This in particular means that the field-mediated interactions between electronic transitions become a second order process. In what follows, we would like to deal with an effective hamiltonian, in which the field mediated interaction between different electronic transitions is treated as a first order term. For that, we follow the same stratgy as reference \cite{Dung2002}, but generalise the formalism for an arbitrarily big set of electronic states. Most of the algebra done here mirrors that of said reference, and we refer the reader there for additional indications on how to perform the derivation. We start from the hamiltonian in a minimal coupling scheme \cite{Buhmann2012}, by adapting the Coulomb gauge and neglecting the ponderomotive interaction term of the field, the minimal coupling hamiltonian can be written as:
\begin{equation}
    \hat{H}=\boxed{ \sum_\lambda \iint d\mathbf{r}\, d\omega\,   \hbar \omega \hat {f}_\lambda^\dagger(\mathbf{r},\omega)\hat {f}_\lambda(\mathbf{r},\omega)}\quad + \quad\boxed{ \frac{\hat{p}^2}{2m} + V(\mathbf{r})}\quad + \quad \boxed{\frac{e}{m} \hat{p}\cdot \hat{A}}
\end{equation}
where we have boxed the terms corresponding respectively to the field hamiltonian ($\hat{H}_F$), a term for the electronic eigenstates ($\hat{H}_e$), and the coupling term ($\hat{H}_{e,F}$). We have introduced the $\hat{f}$ operators, as defined in \cite{Buhmann2012}, which act as elemental field excitations, and follow bosonic conmutation relations: 
\begin{gather}
    [\hat{f}_\lambda(\mathbf{r},\omega),\hat{f}_{\lambda'}(\mathbf{r}',\omega')]=[\hat{f}_\lambda^\dagger (\mathbf{r},\omega),\hat{f}_{\lambda'}^\dagger(\mathbf{r}',\omega')]=\mathbf{0},\\
    [\hat{f}_\lambda(\mathbf{r},\omega),\hat{f}_{\lambda'}^\dagger(\mathbf{r}',\omega')]=\delta_{\lambda,\lambda'}\mathbf{\delta}(\mathbf{r}-\mathbf{r}'){\delta}({\omega}-{\omega}').
\end{gather}
Through the electronic hamiltonian, we obtain the electronic wavefunctions that fulfill $\hat{H}_e\phi_i=E_i\phi_i$, and use these as a basis for the electronic states. As such we can write the electronic term as 
\begin{equation}
    \hat{H}_e=\sum_i \hat{c}_i^{\dagger}\hat{c}_i E_i
\end{equation}
where $\hat{c}_i$ is the annihilation operator of the state $\phi_i$. In the same way, we may introduce these wavefunctions in the interaction term:
\begin{flalign}
    \hat{p}\cdot \hat{A}=&-i \hbar\sum_{i,j}\hat{c}_i^{\dagger}\hat{c}_j\int d\mathbf{r}\, \phi_i^*(\mathbf{r})\nabla\phi_j(\mathbf{r}) \cdot \hat{A}(\mathbf{r})=\nonumber \\
    =&\frac{-i \hbar}{2}\sum_{i,j}\left[ \hat{c}_i^{\dagger}\hat{c}_j\left(\int d\mathbf{r}\, \phi_i^*(\mathbf{r})\nabla\phi_j(\mathbf{r}) \cdot \hat{A}(\mathbf{r})\right)-\left(\int d\mathbf{r}\, \hat{A}(\mathbf{r}) \cdot \phi_i(\mathbf{r})\nabla\phi_j^*(\mathbf{r}) \right)\hat{c}_j^{\dagger}\hat{c}_i\right]
\end{flalign}
where we have expressed the interaction term in a manifestly hermitian way. We define $\hat{\sigma}_{ij}= \hat{c}_i^{\dagger}\hat{c}_j$ and $\mathbf{d}_{ij}(\mathbf{r})=\phi_i^*(\mathbf{r})\nabla\phi_j(\mathbf{r})$. Note that the main difference between this work and reference \cite{Dung2002} is that we are allowing for arbitrary electronic transitions mediated by photons to take place, which leads to the natural apparition of these extended transition dipole densities $\mathbf{d}_{ij}(\mathbf{r})$. This adds complexity in the expressions, but the spirit of the derivation remains unchanged. We now turn to express the vector potential as a function of the $\hat{f}_\lambda$ operators as:
\begin{gather}
    \hat{A}(\mathbf{r})=-i\sum_{\lambda} \int \frac{d\omega}{\omega}\int d\mathbf{r}'\left[\overline{\overline{G}}_\lambda(\mathbf{r},\mathbf{r}',\omega)\cdot \hat{f}_\lambda(\mathbf{r}',\omega)- \hat{f}_\lambda^\dagger(\mathbf{r}',\omega) \cdot \overline{\overline{G}}_\lambda^\dagger(\mathbf{r},\mathbf{r}',\omega) \right]
\end{gather}
Plugging this into the previous Equation and making the aforementioned substitutions leads to
\begin{equation} 
    \begin{split}\label{eq:pAcoupling_def}
       \hat{p}\cdot \hat{A}=\frac{-\hbar}{2}\sum_{i,j}\sum_\lambda\iiint d\mathbf{r}\, d\mathbf{r}'\, \frac{d\omega}{\omega} \Bigg[&\hat{\sigma}_{ij}\, \mathbf{d}_{ij}(\mathbf{r}) \cdot \overline{\overline{G}}_\lambda(\mathbf{r},\mathbf{r}',\omega)\cdot \hat{f}_\lambda(\mathbf{r}',\omega)+\\
       - &\hat{\sigma}_{ij}\, \hat{f}_\lambda^\dagger(\mathbf{r}',\omega) \cdot \overline{\overline{G}}_\lambda^\dagger(\mathbf{r},\mathbf{r}',\omega)\cdot \mathbf{d}_{ij}(\mathbf{r}) \\
       -&\mathbf{d}_{ij}^*(\mathbf{r})\cdot \overline{\overline{G}}_\lambda(\mathbf{r},\mathbf{r}',\omega)\cdot \hat{f}_\lambda(\mathbf{r}',\omega)\, \hat{\sigma}_{ij}^{\dagger}\\ 
       +&\hat{f}_\lambda^\dagger(\mathbf{r}',\omega) \cdot \overline{\overline{G}}_\lambda^\dagger(\mathbf{r},\mathbf{r}',\omega)\cdot \mathbf{d}_{ij}^*(\mathbf{r})\, \hat{\sigma}_{ij}^{\dagger}\Bigg] 
    \end{split}
\end{equation}
In order to obtain direct interaction terms between the different electronic transitions, it is necessary to obtain the dynamics of the $\hat{f}$ operators in terms of the electronic transition operators. The time evolution of any operator in the Heisenberg picture is given by $\dot{\hat{O}}(t)=-\frac{i}{\hbar}[\hat{O}(t),\hat{H}]$. In particular, it is useful to split the complete hamiltonian into the field + electronic system hamiltonian, $\hat{H}_s$, and the interaction hamiltonian $\hat{H}_I=e\, \hat{p}\cdot \hat{A} /m$. This way the evolution of $\hat{O}$ is given by:  
\begin{equation}
\label{eq:Heisemberg_RNR}
    \begin{split}
        \dot{\hat{O}}(t)=-\frac{i}{\hbar}[\hat{O}(t),\hat{H}_s] +& i\frac{e}{2 m}\sum_{i,j}\sum_\lambda\iiint d\mathbf{r}\, d\mathbf{r}'\, \frac{d\omega}{\omega}\times\\
        \Bigg[& [\hat{O},\hat{\sigma}_{ij}] \Bigg( \mathbf{d}_{ij}(\mathbf{r}) \cdot \overline{\overline{G}}_\lambda(\mathbf{r},\mathbf{r}',\omega)\cdot \hat{f}_\lambda(\mathbf{r}',\omega)- \hat{f}_\lambda^\dagger(\mathbf{r}'.\omega) \cdot \overline{\overline{G}}_\lambda^\dagger(\mathbf{r},\mathbf{r}',\omega)\cdot \mathbf{d}_{ij}(\mathbf{r})\Bigg)\\
        &-\Bigg( \mathbf{d}_{ij}^*(\mathbf{r}) \cdot \overline{\overline{G}}_\lambda(\mathbf{r},\mathbf{r}',\omega)\cdot \hat{f}_\lambda(\mathbf{r}',\omega)- \hat{f}_\lambda^\dagger(\mathbf{r}'.\omega) \cdot \overline{\overline{G}}_\lambda^\dagger(\mathbf{r},\mathbf{r}',\omega)\cdot \mathbf{d}_{ij}^*(\mathbf{r})\Bigg) [\hat{O},\hat{\sigma}_{ij}^\dagger]\Bigg]
    \end{split}   
\end{equation}
In particular, the time evolution of the fundamental field operators $\hat{f}_\lambda$ reads:
\begin{flalign*}
     \dot{\hat{f}}_{\lambda'}(\mathbf{r}',\omega',t)=&-\frac{i}{\hbar}\left( \sum_{\lambda}\iint dr d\omega\, \hbar \omega\, \left[\hat{f}_{\lambda'}(\mathbf{r}',\omega',t),\hat{f}_{\lambda}^{\dagger}(\mathbf{r},\omega,t)\right]\hat{f}_{\lambda}(\mathbf{r},\omega,t)+\frac{e}{m}\left[\hat{f}_{\lambda'}(\mathbf{r}',\omega',t),\hat{p}\cdot\hat{A}\right]\right)&&\\
     =&-i\omega'\hat{f}_{\lambda'}(\mathbf{r}',\omega',t) +\frac{ie}{2m\omega'}\sum_{i,j}\int d\mathbf{r}\,     \Bigg[ \overline{\overline{G}}_{\lambda'}^\dagger(\mathbf{r},\mathbf{r}',\omega')\cdot \mathbf{d}_{ij}^*(\mathbf{r})\, \sigma_{ij}^{\dagger}- \sigma_{ij}\,  \overline{\overline{G}}_{\lambda'}^\dagger(\mathbf{r},\mathbf{r}',\omega')\cdot \mathbf{d}_{ij}(\mathbf{r})\Bigg]
\end{flalign*}
Where we have used that $\left[\hat{f}_{\lambda'}(\mathbf{r}',\omega',t),\hat{f}_{\lambda}^{\dagger}(\mathbf{r},\omega,t)\right]=\delta_{\lambda,\lambda'}\delta(\omega-\omega')\mathbf{\delta}(\mathbf{r}-\mathbf{r}')$ and Equation (\ref{eq:pAcoupling_def}). Solving this equation formally we find 
\begin{flalign*}
     \hat{f}_{\lambda'}(\mathbf{r}',\omega,t)=&\hat{f}_{\lambda'}^{\mbox{free}}(\mathbf{r}',\omega,t)-i \frac{e}{2m\omega}\sum_{k,l}\times&&\\
     &\left\{ \iint_0^t dt'\, d\mathbf{r}''\,\left[ \sigma_{kl}(t')  \overline{\overline{G}}_{\lambda'}^\dagger(\mathbf{r}'',\mathbf{r}',\omega)\cdot \mathbf{d}_{kl}(\mathbf{r}'') - \overline{\overline{G}}_{\lambda'}^\dagger(\mathbf{r}'',\mathbf{r}',\omega)\cdot \mathbf{d}_{kl}^*(\mathbf{r}'')\, \sigma_{kl}^{\dagger}(t')\right]e^{-i\omega (t-t')}\right\}\\
\end{flalign*}
plugging this back into Equation (\ref{eq:Heisemberg_RNR}), expanding the product and making use of the identity $\sum_\lambda \int d\mathbf{s} \overline{\overline{G}}_\lambda(\mathbf{r},\mathbf{s},\omega)\cdot \overline{\overline{G}}_{\lambda}^\dagger(\mathbf{r}'',\mathbf{s},\omega)=\frac{\hbar \mu_0}{\pi}\omega^2 \mbox{Im}(\overline{\overline{G}}(\mathbf{r},\mathbf{r}'',\omega))$ \cite{Buhmann2012}, the equation of motion for $\hat{O}$ reads
\begin{flalign*}
        \dot{\hat{O}}(t)=-\frac{i}{\hbar}[\hat{O}(t),\hat{H}_s] &+ i\frac{e^2}{4 m^2}\frac{\hbar \mu_0}{\pi}\sum_{\underset{k,l}{i,j}}\int_0^t dt'\iiint d\mathbf{r}\,  d\mathbf{r}''\,  d\omega\times\\
        \Bigg[& -i[\hat{O},\sigma_{ij}] \sigma_{kl}(t') \mathbf{d}_{ij}(\mathbf{r}) \cdot \mbox{Im}\left\{\overline{\overline{G}}(\mathbf{r},\mathbf{r}'',\omega)\right\}\cdot \mathbf{d}_{kl}(\mathbf{r}'') e^{-i\omega (t-t')}\\
        &+i[\hat{O},\sigma_{ij}] \sigma_{kl}^{\dagger}(t') \mathbf{d}_{ij}(\mathbf{r}) \cdot \mbox{Im}\left\{\overline{\overline{G}}(\mathbf{r},\mathbf{r}'',\omega)\right\}\cdot \mathbf{d}_{kl}^*(\mathbf{r}'')\, e^{-i\omega (t-t')}\\
        &-  i[\hat{O},\sigma_{ij}] \sigma_{kl}^\dagger(t')\, \mathbf{d}^*_{kl}(\mathbf{r}'')\cdot \mbox{Im}\left\{\overline{\overline{G}}(\mathbf{r}'',\mathbf{r},\omega)\right\}\cdot \mathbf{d}_{ij}(\mathbf{r}) e^{i\omega (t-t')}\\
        &+  i[\hat{O},\sigma_{ij}] \sigma_{kl}(t')  \mathbf{d}_{kl}(\mathbf{r}'')\cdot\mbox{Im}\left\{\overline{\overline{G}}(\mathbf{r}'',\mathbf{r},\omega)\right\}\cdot \mathbf{d}_{ij}(\mathbf{r}) e^{i\omega (t-t')}\\
        & +i\mathbf{d}_{ij}^*(\mathbf{r}) \cdot \mbox{Im}\left\{\overline{\overline{G}}(\mathbf{r},\mathbf{r}'',\omega)\right\}\cdot \mathbf{d}_{kl}(\mathbf{r}'') \sigma_{kl}(t') [\hat{O},\sigma_{ij}^\dagger]e^{-i\omega (t-t')}\\
        &-i\mathbf{d}_{ij}^*(\mathbf{r}) \cdot \mbox{Im}\left\{\overline{\overline{G}}(\mathbf{r},\mathbf{r}'',\omega)\right\}\cdot \mathbf{d}_{kl}^*(\mathbf{r}'')\, \sigma_{kl}^{\dagger}(t') [\hat{O},\sigma_{ij}^\dagger]e^{-i\omega (t-t')}\\
        &+ i\mathbf{d}^*_{kl}(\mathbf{r}'')\cdot \mbox{Im}\left\{\overline{\overline{G}}(\mathbf{r}'',\mathbf{r},\omega)\right\}\cdot \mathbf{d}_{ij}^*(\mathbf{r})  \sigma_{kl}^\dagger(t') [\hat{O},\sigma_{ij}^\dagger]e^{i\omega (t-t')}\\
        &-i \mathbf{d}_{kl}(\mathbf{r}'')\mbox{Im}\left\{\overline{\overline{G}}(\mathbf{r}'',\mathbf{r},\omega)\right\}\cdot \mathbf{d}_{ij}^*(\mathbf{r})\, \sigma_{kl}(t')  [\hat{O},\sigma_{ij}^\dagger] e^{i\omega (t-t')}\Bigg]
\end{flalign*}
Now we introduce a coarse-graining Markov approximation, approximating the time dependency of the ladder operators by the natural frequency of the shifted eigenenergies\footnote{Solving a Heisemberg equation for the time evolution of these ladder operators one can see that $\omega_{ij}=\omega_j - \omega_i$.}: $\hat{\sigma}_{ij}(t')\approx \tilde{\hat{\sigma}}_{ij}(t) e^{-i\omega_{ij}t'}$. This in turn means that $\tilde{\hat{\sigma}}_{ij}(t)=\hat{\sigma}_{ij}(t)e^{i\omega_{ij}t} $, and therefore $\hat{\sigma}_{ij}(t')\approx \hat{\sigma}_{ij}(t) e^{i\omega_{ij}(t-t')}$. Then we make the substitution $\int dt' e^{i\Omega(t-t')}\rightarrow \xi(\Omega)$\label{eq:Sokhotski-Plemejl_decomp}, where $\xi(x)=\pi \delta(x)+i \mathcal{P}\left( 1/x \right)$ by virtue of the Sokhotski-Plemejl theorem, where $\mathcal{P}$ denotes the principal value. These two terms of the integral represent the resonant (the $\delta$ function) and the non resonant (the principal value) part of the interaction. In what follows, we assume that the field-mediated interaction between the two electronic transitions is happening through the non-resonant electromagnetic environment, and therefore only consider the principal value contribution. In this situation $\xi(-x)=\xi^*(x)$, and the time evolution of $\hat{O}$ reads: 
\begin{flalign*}
        \dot{\hat{O}}(t)=-\frac{i}{\hbar}[\hat{O}(t),\hat{H}_s] &+ i\frac{e^2}{4 m^2}\frac{\hbar \mu_0}{\pi}\sum_{\underset{k,l}{i,j}}\iiint d\mathbf{r}\,  d\mathbf{r}''\,  d\omega\times\\
        \Bigg[& -i[\hat{O},\sigma_{ij}] \sigma_{kl} \mathbf{d}_{ij}(\mathbf{r}) \cdot \mbox{Im}\left\{\overline{\overline{G}}(\mathbf{r},\mathbf{r}'',\omega)\right\}\cdot \mathbf{d}_{kl}(\mathbf{r}'') \xi(-(\omega-\omega_{kl}))\\
        &+i[\hat{O},\sigma_{ij}] \sigma_{kl}^{\dagger} \mathbf{d}_{ij}(\mathbf{r}) \cdot \mbox{Im}\left\{\overline{\overline{G}}(\mathbf{r},\mathbf{r}'',\omega)\right\}\cdot \mathbf{d}_{kl}^*(\mathbf{r}'')\, \xi(-(\omega+\omega_{kl}))\\
        &-  i[\hat{O},\sigma_{ij}] \sigma_{kl}^\dagger\, \mathbf{d}^*_{kl}(\mathbf{r}'')\cdot \mbox{Im}\left\{\overline{\overline{G}}(\mathbf{r}'',\mathbf{r},\omega)\right\}\cdot \mathbf{d}_{ij}(\mathbf{r}) \xi(\omega-\omega_{kl})\\
        &+  i[\hat{O},\sigma_{ij}] \sigma_{kl}  \mathbf{d}_{kl}(\mathbf{r}'')\cdot\mbox{Im}\left\{\overline{\overline{G}}(\mathbf{r}'',\mathbf{r},\omega)\right\}\cdot \mathbf{d}_{ij}(\mathbf{r}) \xi(\omega+\omega_{kl})\\
        & +i\mathbf{d}_{ij}^*(\mathbf{r}) \cdot \mbox{Im}\left\{\overline{\overline{G}}(\mathbf{r},\mathbf{r}'',\omega)\right\}\cdot \mathbf{d}_{kl}(\mathbf{r}'') \sigma_{kl} [\hat{O},\sigma_{ij}^\dagger]\xi(-(\omega-\omega_{kl}))\\
        &-i\mathbf{d}_{ij}^*(\mathbf{r}) \cdot \mbox{Im}\left\{\overline{\overline{G}}(\mathbf{r},\mathbf{r}'',\omega)\right\}\cdot \mathbf{d}_{kl}^*(\mathbf{r}'')\, \sigma_{kl}^{\dagger} [\hat{O},\sigma_{ij}^\dagger]\xi(-(\omega+\omega_{kl}))\\
        &+ i\mathbf{d}^*_{kl}(\mathbf{r}'')\cdot \mbox{Im}\left\{\overline{\overline{G}}(\mathbf{r}'',\mathbf{r},\omega)\right\}\cdot \mathbf{d}_{ij}^*(\mathbf{r})  \sigma_{kl}^\dagger [\hat{O},\sigma_{ij}^\dagger]\xi(\omega-\omega_{kl})\\
        &-i \mathbf{d}_{kl}(\mathbf{r}'')\mbox{Im}\left\{\overline{\overline{G}}(\mathbf{r}'',\mathbf{r},\omega)\right\}\cdot \mathbf{d}_{ij}^*(\mathbf{r})\, \sigma_{kl}  [\hat{O},\sigma_{ij}^\dagger] \xi(\omega+\omega_{kl})\Bigg]
\end{flalign*}
Using several of the exchange symmetries of the indices, namely $d_{ji}=-d_{ij}^*$ and that $\sigma_{kl}=\sigma_{lk}^\dagger$, and $\omega_{lk}=-\omega_{kl}$, allows to rewrite the previous Heisenberg equation in a more compact form
\begin{flalign} 
        \dot{\hat{O}}(t)=&-\frac{i}{\hbar}[\hat{O}(t),\hat{H}_s] + \frac{i}{\hbar}\frac{e^2}{2 m^2}\frac{\hbar^2 \mu_0}{\pi}\sum_{\substack{i,j \\ k,l}}
        \Bigg[[\hat{O},\sigma_{ij}]\sigma_{kl}^{\dagger} f_{ij,kl}+f_{ij,kl}^*\sigma_{kl}[\hat{O},\sigma_{ij}^\dagger] \Bigg]
\end{flalign}
where we have defined:
\begin{flalign} 
        f_{ij,kl}=&\iint d\mathbf{r}\, d\mathbf{r}'\,\mathbf{d}_{ij}(\mathbf{r}) \cdot\left\{  \mathcal{P}\int_0^\infty d\omega\left[\frac{ \mbox{Im}\left\{\overline{\overline{G}}(\mathbf{r},\mathbf{r}',\omega)\right\}}{\omega_{kl}+\omega}+\frac{\mbox{Im}\left\{\overline{\overline{G}}(\mathbf{r},\mathbf{r}',\omega)\right\}}{\omega-\omega_{kl}} \right]\right\} \cdot \mathbf{d}_{kl}^*(\mathbf{r}')
\end{flalign}
In these coefficients, we have integrals regarding the imaginary part of the system greens function. We can use the kramers-kronig relations to perform these integrals in the following way:
\begin{gather*}
    \mathcal{P} \int_0^\infty d\omega\, \frac{ \mbox{Im}\left\{\overline{\overline{G}}(\mathbf{s},\mathbf{t},\omega)\right\}}{\omega \pm \omega_{0}} =\mathcal{P} \int_{-\infty}^\infty d\omega\, \frac{ \mbox{Im}\left\{\overline{\overline{G}}(\mathbf{s},\mathbf{t},\omega)\right\}}{\omega \pm \omega_{0}} -\mathcal{P} \int_{-\infty}^0 d\omega\, \frac{ \mbox{Im}\left\{\overline{\overline{G}}(\mathbf{s},\mathbf{t},\omega)\right\}}{\omega \pm \omega_{0}}=\\
    =\pi \mbox{Re}\left\{\overline{\overline{G}}(\mathbf{s},\mathbf{t},\mp\omega_{0})\right\}-\mathcal{P} \int_{0}^\infty d\omega\, \frac{ \mbox{Im}\left\{\overline{\overline{G}}(\mathbf{s},\mathbf{t},\omega)\right\}}{\omega \mp \omega_{0}}
\end{gather*}
Where we have used the following property of the greens function $\overline{\overline{G}}^*(\mathbf{s},\mathbf{t},\omega)=\overline{\overline{G}}(\mathbf{s},\mathbf{t},-\omega^*)$. Taking into account that the Dyadic is a symmetric tensor and that the spatial coordinates can be exchanged \cite{Novotny2009}, together with the relation we have just shown, we can further simplify the constants as
\begin{flalign*} 
        f_{ij,kl}=\pi \iint d\mathbf{r}\, d\mathbf{r}'\,\mathbf{d}_{ij}(\mathbf{r}) \cdot \mbox{Re}\left\{\overline{\overline{G}}(\mathbf{r},\mathbf{r}',|\omega_{kl}|)\right\} \cdot \mathbf{d}_{kl}^*(\mathbf{r}').
\end{flalign*}
By further manipulating the master equation it can be cast in the following form:
\begin{flalign} 
        \dot{\hat{O}}(t)=&-\frac{i}{\hbar}[\hat{O}(t),\hat{H}_s] \nonumber \\ \label{eq:MasterAlmostdone}
        &+ \frac{i}{\hbar}\frac{e^2}{2 m^2}\frac{\hbar^2 \mu_0}{\pi}\sum_{\substack{i,j \\ k,l}}
        \Bigg[\frac{f_{ij,kl}+f_{kl,ij}^*}{2}\left[\hat{O},\sigma_{ij}\sigma_{kl}^{\dagger}\right]+\frac{f_{ij,kl}-f_{kl,ij}^*}{2}\left[\left\{\hat{O},\sigma_{ij}\sigma_{kl}^{\dagger}\right\}-2\sigma_{ij}\hat{O}\sigma_{kl}^\dagger\right]\Bigg]
\end{flalign}
we recognize two terms of different nature: the first one is hermitian, and captures a coherent coupling between the two electronic transitions mediated by the EM fields. The second term has a lindblad-like form, and describes the dissipative coupling between these two transitions. Explicitly writing the coefficients of these two terms:
\begin{gather*}
f_{ij,kl}+f_{kl,ij}^*=\pi \iint d\mathbf{r}\, d\mathbf{r}'\,\mathbf{d}_{ij}(\mathbf{r}) \cdot \left[ \mbox{Re}\left\{\overline{\overline{G}}(\mathbf{r},\mathbf{r}',|\omega_{kl}|)\right\}+\mbox{Re}\left\{\overline{\overline{G}}(\mathbf{r},\mathbf{r}',|\omega_{ij}|)\right\}  \right]\cdot \mathbf{d}_{kl}^*(\mathbf{r}'),\\
    f_{ij,kl}-f_{kl,ij}^*=\pi \iint d\mathbf{r}\, d\mathbf{r}'\,\mathbf{d}_{ij}(\mathbf{r}) \cdot \left[ \mbox{Re}\left\{\overline{\overline{G}}(\mathbf{r},\mathbf{r}',|\omega_{kl}|)\right\}-\mbox{Re}\left\{\overline{\overline{G}}(\mathbf{r},\mathbf{r}',|\omega_{ij}|)\right\}  \right]\cdot \mathbf{d}_{kl}^*(\mathbf{r}'),
\end{gather*}
one sees that if the two electronic transitions involved are resonant or happen in frequencies in which the Green's function varies slowly, then $f_{ij,kl}\approx f_{kl,ij}^*$, and then the dynamics of the field mediated electronic interaction will be described by an effective hamiltonian $\dot{\hat{O}}(t)=-\frac{i}{\hbar}[\hat{O}(t),\hat{H}_{eff}]$ and the hamiltonian is defined as:
\begin{gather} 
        \hat{H}_{eff}=\hat{H}_s - \sum_{ij}\hbar  g_{ij,ij}^{m-m} \hat{c}_i^\dagger \hat{c}_i - \sum_{\substack{i,j \\ k,l\neq i,j }}\hbar  g_{ij,kl}^{m-m} \sigma_{ij}\sigma_{kl}^{\dagger} \label{eq:eff_Ham_almost_ready}\\
        g_{ij,kl}^{m-m}=\frac{e^2 \hbar \mu_0}{2 m^2}\iint d\mathbf{r}\, d\mathbf{r}'\,\mathbf{d}_{ij}(\mathbf{r}) \cdot \mbox{Re}\left\{\overline{\overline{G}}\left(\mathbf{r},\mathbf{r}',\frac{|\omega_{kl}|+|\omega_{ij}|}{2}\right)\right\}\cdot \mathbf{d}_{kl}^*(\mathbf{r}')
\end{gather}
Where we have introduced an effective matter-matter coupling strength $g_{ij,kl}^{m-m}$. Note that in \Cref{eq:eff_Ham_almost_ready} we distinguish between cases in which $\{kl\}\neq\{ij\}$, and those in which $\{kl\}=\{ij\}$. While the first summation terms describe this coherent coupling between electronic transitions, the second kind renormalizes the energy of the electronic eigenstates with an energy shift given by $\hbar\delta_i=\sum_j g_{ij,ij}^{m-m}$. From the definition of the coupling strength we see that $g_{ij,kl}^{m-m}={g_{kl,ij}^{m-m}}^*$, and also, since $\mathbf{d}_{ji}=-\mathbf{d}_{ij}^*$, $g_{ji,lk}^{m-m}={g_{ij,kl}^{m-m}}^*$. The complete hamiltonian then reads:
\begin{flalign}  \label{eq:effectiveHamiltonian} 
    \hat{H}_{\mbox{eff}}=& \sum_\lambda \iint  d\mathbf{r}\, d\omega\,   \hbar \omega \hat {f}_\lambda^\dagger(\mathbf{r},\omega)\hat {f}_\lambda(\mathbf{r},\omega)
    + \sum_i \left(E_i - \hbar\delta_i \right)\hat{c}_i^\dagger \hat{c}_i\nonumber \\
    +&\frac{e}{m} \hat{p}\cdot \hat{A}'
    -\sum_{\substack{i,j \\ k,l\neq i,j }}\hbar\,
    g_{ij,kl}^{m-m}\,\sigma_{ij}\sigma_{kl}^{\dagger}
\end{flalign}
The first two terms in \Cref{eq:effectiveHamiltonian} correspond to the energies of the field and electronic states, while the second line describes the coupling between the field and the electronic states and among electronic transitions. It is important to notice that the interaction of the electronic states with the field is mediated by the resonant frequency part, which we did not account for in the derivation of the last term\footnote{See the Sokhotski-Plemejl decomposition made in page \pageref{eq:Sokhotski-Plemejl_decomp} }. On the next section, we will focus on writing the light-matter coupling term of the hamiltonian in a similar fashion as we have done here.

\subsection{Light-matter interaction: Current centered modes} \label{appendix:light-matter_interaction_hamiltonian_a_la_Johannes}
Next, our interest lies in finding how to express the light matter couplings in terms of generalized bosonic operators. For that we revisit the resonant term in the minimal coupling in Equation (\ref{eq:pAcoupling_def}). Writing:
\begin{flalign*}
\hat{p}=&-i \frac{\hbar}{2}\sum_{i,j}\int d\mathbf{r} \left[\hat{\sigma}_{ij}\, \phi_i^*(\mathbf{r})\nabla\phi_j(\mathbf{r})-\hat{\sigma}_{ji} \, \phi_i(\mathbf{r})\nabla\phi_j^*(\mathbf{r}) \right]\\
\hat{A}(\mathbf{r})=&-i\sum_{\lambda} \int \frac{d\omega}{\omega}\int d\mathbf{r}'\left[\overline{\overline{G}}_\lambda(\mathbf{r},\mathbf{r}',\omega)\cdot \hat{f}_\lambda(\mathbf{r}',\omega)- \hat{f}_\lambda^\dagger(\mathbf{r}'.\omega) \cdot \overline{\overline{G}}_\lambda^\dagger(\mathbf{r},\mathbf{r}',\omega) \right]
\end{flalign*}
The interaction hamiltonian term looks like
\begin{flalign}
       H_I=\frac{e}{m}\hat{p}\cdot \hat{A}=-\frac{e \hbar}{2m}\sum_{i,j} \sum_{\lambda}\int \frac{d\omega}{\omega} \Bigg\{& \hat{\sigma}_{ij} \iint d\mathbf{r}\, d\mathbf{r}'\, \mathbf{d}_{ij}(\mathbf{r}) \cdot \overline{\overline{G}}_\lambda(\mathbf{r},\mathbf{r}',\omega)\cdot \hat{f}_\lambda(\mathbf{r}',\omega)\nonumber\\
       +& \hat{\sigma}_{ji} \iint d\mathbf{r}\, d\mathbf{r}'\, \mathbf{d}_{ji}(\mathbf{r}) \cdot \overline{\overline{G}}_\lambda(\mathbf{r},\mathbf{r}',\omega)\cdot \hat{f}_\lambda(\mathbf{r}',\omega)\nonumber\\
       +&\hat{\sigma}_{ij}  \iint d\mathbf{r}\, d\mathbf{r}'\, \hat{f}_\lambda^\dagger(\mathbf{r}',\omega) \cdot \overline{\overline{G}}_\lambda^\dagger(\mathbf{r},\mathbf{r}',\omega) \cdot \mathbf{d}_{ji}^*(\mathbf{r})\nonumber\\
       +&\hat{\sigma}_{ji}  \iint d\mathbf{r}\, d\mathbf{r}'\, \hat{f}_\lambda^\dagger(\mathbf{r}',\omega) \cdot \overline{\overline{G}}_\lambda^\dagger(\mathbf{r},\mathbf{r}',\omega) \cdot  \mathbf{d}_{ij}^*(\mathbf{r})\Bigg\}
\end{flalign}
 where to get the last term we have just exchanged the indices and used that $\mathbf{d}_{ji}(\mathbf{r})=-\mathbf{d}_{ij}^*(\mathbf{r})$. Following the spirit of emitter centerd modes \cite{Feist2020,Buhmann2008} we rewrite the hamiltonian by defining a new set of bosonic operators as 
\begin{flalign}
       H_I&=\hbar\sum_{i,j}\int d\omega \Bigg[g_{ij}(\omega) \sigma_{ij}\,\hat{a}_{ij}(\omega)+g_{ji}(\omega) \sigma_{ji}\,\hat{a}_{ji}(\omega) +g_{ij}^*(\omega) \sigma_{ij}^\dagger\,\hat{a}_{ij}^\dagger(\omega)+g_{ji}^*(\omega) \sigma_{ji}^\dagger\,\hat{a}_{ji}^\dagger(\omega) \Bigg]\label{eq:Hint_reduced}\\
       \hat{a}_{ij}(\omega)&=\frac{-e\sum_\lambda}{2m\omega g_{ij}(\omega)} \iint d\mathbf{r}\, d\mathbf{r}'\, \mathbf{d}_{ij}(\mathbf{r}) \cdot \overline{\overline{G}}_\lambda(\mathbf{r},\mathbf{r}',\omega)\cdot \hat{f}_\lambda(\mathbf{r}',\omega) \label{eq:currentcentered_a}\\
       \hat{a}_{ij}^\dagger(\omega)&=\frac{-e\sum_\lambda}{2m \omega g_{ij}^*(\omega)} \iint d\mathbf{r}\, d\mathbf{r}'\, \hat{f}_\lambda^\dagger(\mathbf{r}',\omega) \cdot \overline{\overline{G}}_\lambda^\dagger(\mathbf{r},\mathbf{r}',\omega)\cdot \mathbf{d}_{ij}^{*}(\mathbf{r})
\end{flalign}
From these expressions and the conmutation properties of the $\hat{f}$ operators, it can be shown that $[\hat{a}_{ij},\hat{a}_{kl}]=0$. By using the conmutation relations of the $\hat{f}$ operators and the identity $\sum_\lambda \int ds \overline{\overline{G}}_\lambda(\mathbf{r},\mathbf{s},\omega)\cdot \overline{\overline{G}}_{\lambda}^\dagger(\mathbf{r}'',\mathbf{s},\omega)=\frac{\hbar \mu_0}{\pi}\omega^2 \mbox{Im}(\overline{\overline{G}}(\mathbf{r},\mathbf{r}'',\omega))$, the conmutation relation for these new bosonic operators is given by:
\begin{flalign*}
    [\hat{a}_{ij}(\omega),\hat{a}_{kl}^\dagger(\omega')]=\delta(\omega-\omega')\frac{e^2\hbar \mu_0}{4m^2\pi g_{ij}(\omega) g_{kl}^*(\omega)}\iint d\mathbf{r} d\mathbf{r}''\,  \mathbf{d}_{ij}(\mathbf{r})\cdot \Im\Big\{\overline{\overline{G}}(\mathbf{r},\mathbf{r}'',\omega)\Big\} \cdot \mathbf{d}_{kl}^*(\mathbf{r}'') 
\end{flalign*}
In particular, by imposing $[\hat{a}_{ij},\hat{a}_{ij}^\dagger]\equiv 1$, we find that the coupling between the photon mode and the electronic transition is given by: 
\begin{gather}
    g_{ij}(\omega)=\frac{e}{2m}\sqrt{\frac{\hbar \mu_0}{\pi}\iint d\mathbf{r}\, d\mathbf{r}'\, \mathbf{d}_{ij}(\mathbf{r})\cdot \Im\Big\{\overline{\overline{G}}(\mathbf{r},\mathbf{r}',\omega)\Big\} \cdot \mathbf{d}_{ij}^*(\mathbf{r}')} \label{eq:coupling_strength}
\end{gather}
This is the central result of this section. It shows that the coupling strength between a given electronic transition and an optical mode can be written in a form very similar to the coupling strength between a QE and an optical mode, albeit replacing the puntual dipole moment by an extended current distribution. Using the fact that the greens function is a symmetric tensor, it can be seen that $g_{ij}$ is a real constant, and that $g_{ij}=g_{ji}$. This then allows to rewrite the general conmutator as
\begin{flalign}
\label{eq:Conmutator_current_centered_modes}
    &[\hat{a}_{ij}(\omega),\hat{a}_{kl}^\dagger(\omega')]=\delta(\omega-\omega')\frac{F_{ij}^{kl}}{\sqrt{F_{ij}^{ij} F_{kl}^{kl}}}\\
    &F_{ab}^{cd}=\iint d\mathbf{r} d\mathbf{r}' \, \mathbf{d}_{ab}(\mathbf{r})\cdot \Im\Big\{\overline{\overline{G}}(\mathbf{r},\mathbf{r}',\omega)\Big\} \cdot \mathbf{d}_{cd}^*(\mathbf{r}')
\end{flalign}
Although we will not tackle this problem here, the modes defined in this manner are not orthogonal, since the conmutators for the modes from different transitions are non-zero. Some strategies have been proposed to tackle this problem and generate a set of field operators that do conmute among themselves \cite{Feist2020}. We are particularly interested in systems in which the Green's funciton is sharply peaked around a resonant frequency: $\omega_c$. By writting the Greens function as 
\begin{gather}
\overline{\overline{G}}(\mathbf{r},\mathbf{r}',\omega)=\overline{\overline{\mathcal{G}}}(\mathbf{r},\mathbf{r}',\omega_c)\delta(\omega-\omega_c),   \label{eq:resonant_GF_aux} 
\end{gather}
the interaction hamiltonian in \Cref{eq:Hint_reduced} can be written as a familiar Rabi interaction hamiltonian:
\begin{flalign}
       H_I&=\hbar\sum_{i,j}g_{ij}^{l-m}\Bigg[ \sigma_{ij} \left(\hat{a}_{ij}+\hat{a}_{ji}^\dagger\right)+ \sigma_{ij}^\dagger\left(\hat{a}_{ji} +\hat{a}_{ij}^\dagger\right) \Bigg]\label{eq:RabiMQED_intHam}\\
       \hat{a}_{ij}&=\frac{-e\sum_\lambda}{2m\omega g_{ij}} \iint d\mathbf{r}\, d\mathbf{r}'\, \mathbf{d}_{ij}(\mathbf{r}) \cdot \overline{\overline{\mathcal{G}}}_\lambda(\mathbf{r},\mathbf{r}',\omega_c)\cdot \hat{f}_\lambda(\mathbf{r}',\omega_c) \label{eq:currentcentered_a_single_freq}\\
       g_{ij}^{l-m}&=\frac{e}{2m}\sqrt{\frac{\hbar \mu_0}{\pi}\iint d\mathbf{r}\, d\mathbf{r}'\, \mathbf{d}_{ij}(\mathbf{r})\cdot \Im\Big\{\overline{\overline{\mathcal{G}}}(\mathbf{r},\mathbf{r}',\omega_c)\Big\} \cdot \mathbf{d}_{ij}^*(\mathbf{r}')} \label{eq:coupling_strength_single_freq}
\end{flalign}
Where we have now introduced a light-matter interaction strength denoted by $g_{ij}^{l-m}$. Finally, the effective hamiltonian of a system that involves general electronic transitions interacting with a single mode field can be written as
\begin{flalign}  
    \hat{H}_{\mbox{eff}}=& \sum_{ij} \hbar \omega_c \hat {a}_{ij}^\dagger \hat{a}_{ij}
    + \sum_i \tilde{E}_i\hat{c}_i^\dagger \hat{c}_i\nonumber \\
    +&\hbar\sum_{i,j}g_{ij}^{l-m} \hat{\sigma}_{ij} \left(\hat{a}_{ij}+\hat{a}_{ji}^\dagger\right) \nonumber\\
    -&\sum_{\substack{i,j \\ k,l\neq i,j}}\hbar\,
    g_{ij,kl}^{m-m}\,\sigma_{ij}\sigma_{kl}^{\dagger} \label{eq:effectiveHamiltonian_complete} 
\end{flalign}
with 
\begin{flalign}
&g_{ij}^{l-m}=\frac{e}{m}\sqrt{\frac{\hbar \mu_0}{\pi}\iint d\mathbf{r}\, d\mathbf{r}'\, \mathbf{d}_{ij}(\mathbf{r})\cdot \Im\Big\{\overline{\overline{\mathcal{G}}}(\mathbf{r},\mathbf{r}',\omega_c)\Big\} \cdot \mathbf{d}_{ij}^*(\mathbf{r}')}\label{eq:general_couplings_lightmatter}\\
&g_{ij,kl}^{m-m}=\frac{e^2 \hbar \mu_0}{2 m^2}\iint d\mathbf{r}\, d\mathbf{r}'\,\mathbf{d}_{ij}(\mathbf{r}) \cdot \mbox{Re}\left\{\overline{\overline{G}}\left(\mathbf{r},\mathbf{r}',\frac{|\omega_{kl}|+|\omega_{ij}|}{2}\right)\right\}\cdot \mathbf{d}_{kl}^*(\mathbf{r}') \label{eq:general_couplings_mattermatter}
\end{flalign}
\section{Electronic transitions of interest} 
\label{SM2:electronic_transitions_of_interest}
One of the strengths of the above hamiltonian is it's applicability to a wide variety of electronic transitions. One could study the coupling of bloch electrons in materials, localised electronic transitions in molecules, or bound electrons in material defects on equal footing, provided one is able of obtaining the initial and final electronic states involved in the transition. In this section, we will apply the above formalism to derive couplings for specific cases: localized dipolar electronic transitions and also free electrons.

\subsection{Localized transitions: dipolar quantum emitters}
One of the most relevant applications of this framework is when dealing with localised transitions. In this family we can include transitions inside quantum dots, molecules or defects in crystals. This is due to the fact that regardless of the actual shape of the electronic states taking place in the transition, the scale lengths associated with them is much smaller than the scale in which the EM fields change. As such, one could assume that the Green's function is constant over the span of the electronic wavefunctions of the quantum emitter (QE) and write: 
\begin{gather}
    \mathbf{d}_{ij}^{\scriptstyle{QE}}(r)\approx \delta(r-r_0)\int d\mathbf{r}'  \mathbf{d}_{ij}(r')=\delta(r-r_0)\int d\mathbf{r}'  \phi_i^*(r')\nabla\phi_j(r')=\delta(r-r_0)\bra{i} \nabla\ket{j}=\nonumber\\
    =-\delta(r-r_0)\frac{m}{\hbar^2}\bra{i} [\hat{H}_e,\hat{r}]\ket{j} =-\delta(r-r_0)\frac{m}{e\hbar} (\omega_j- \omega_i)\bra{i}-e\hat{r}\ket{j}
    =-\delta(r-r_0)\frac{m\omega_{ij}}{e\hbar} \boldsymbol{\mu}_{ij} \label{eq:transition_moment_dipole_transition}
\end{gather}
where in the end we have the energy difference between the initial and final state given by $\omega_{ij}=\omega_j-\omega_i$, and the usual transition dipole moment given by $\boldsymbol{\mu}_{ij}=\bra{i}-e\hat{r}\ket{j}$, where e is the electron charge. To arrive at this expression we have used that, in general, the electronic hamiltonian will be written as $\hat{H}_e= V(r) + \hat{p}^2/2m$ and therefore $[\hat{H}_e,\hat{r}]=-i\hbar \hat{p}/m = -\hbar^2 \nabla /m$. Putting this transition current density into \Cref{eq:general_couplings_lightmatter}, we get the very familiar expression for the coupling between a dipolar QE and an optical mode of a cavity:
\begin{flalign}
g_{ij}^{c-QE}=\frac{\left\lvert \omega_{ij}\right\rvert}{\hbar}\sqrt{\frac{ \hbar\mu_0}{\pi}\boldsymbol{\mu}_{ij}\cdot \Im\Big\{\overline{\overline{\mathcal{G}}}(\mathbf{r}_0,\mathbf{r}_0,\omega_c)\Big\} \cdot \boldsymbol{\mu}_{ij}^*}
\end{flalign}
Since these couplings are real, $g_{ij}^{c-QE}=g_{ji}^{c-QE}\equiv g^{c-QE}$. For the case of a single QE, with eigenstates named by $\ket{g}$ and $\ket{e}$, for ground and excited state, the interaction between this QE and some EM environment is given (from \Cref{eq:effectiveHamiltonian_complete}) by:
\begin{flalign} 
    \hat{H}_{I}^{c-QE}=&g^{c-QE} \hbar\left[ \hat{\sigma}_{ge} \left(\hat{a}_{ge}+\hat{a}_{eg}^\dagger\right) +\hat{\sigma}_{eg} \left(\hat{a}_{eg}+\hat{a}_{ge}^\dagger\right)\right]\nonumber\\
    \approx &\hbar g^{c-QE} \left[  \hat{\sigma} \hat{a}_{eg}^\dagger + \hat{\sigma}^\dagger\hat{a}_{eg}\right] \label{eq:MQED_lm_dip_doupling_gemeral_Hamiltonian}\\
    g^{c-QE}&=\frac{\left\lvert \omega_{QE}\right\rvert}{\hbar}\sqrt{\frac{ \hbar\mu_0}{\pi}\boldsymbol{\mu}\cdot \Im\Big\{\overline{\overline{\mathcal{G}}}(\mathbf{r}_0,\mathbf{r}_0,\omega_c)\Big\} \cdot \boldsymbol{\mu}^*} \label{eq:MQED_lm_dip_doupling_gemeral}
\end{flalign}
where we have introduced the usual QE ladder operators given by $\hat{\sigma}=\dyad{g}{e}$, defined the QE transition dipole moment as $\boldsymbol{\mu}\equiv \boldsymbol{\mu}_{ge}=\bra{g}-e\hat{r}\ket{e}$, and applied the rotating wave approximation to the cavity-QE coupling, which leads to the interaction being written in terms of a single optical mode.

\subsection{Free electrons}
We will approximate free electron states as momentum eigenstates in the $\hat{z}$ direction with some lateral distribution: $\phi_k(\mathbf{r})=g_{\perp}(\mathbf{r}_\perp)e^{ik_k z}/\sqrt{L}$ where $L$ is the length of a fictitious box used to quantize the electron momenta, and $g_{\perp}(\mathbf{r}_\perp)$ is a function describing the thin lateral profile of the electron wavepacket. In principle we could say that 
\begin{gather*}
g_{\perp}(\mathbf{r}_\perp)=\sqrt{\frac{1}{2\pi \sigma^2 }}\exp{-\left(\frac{|\mathbf{r}_\perp -\mathbf{r}_{0,\perp}|}{2\sigma}\right)^2},
\end{gather*}
which in the limit of small beam width behaves as a delta in the lateral direction. For a transition between free states we then have:
\begin{flalign}
 \mathbf{d}_{kl}(\mathbf{r})= \phi_k^*\nabla\phi_l &= \frac{1}{\sigma} \begin{pmatrix} \frac{(x_0-x)}{2\sigma}\\\frac{(y_0-y)}{2\sigma}\\ik_l\sigma \end{pmatrix}\frac{e^{-\frac{|\mathbf{r}_\perp -\mathbf{r}_{0,\perp}|^2}{2\sigma^2}}}{2\pi\sigma^2}\frac{e^{i(k_l-k_k)z}}{L} \label{eq:d_freest_relation_finite_sigma}\\
     &\overset{\lim_{\sigma\rightarrow 0}}{\approx} i k_0 \frac{\delta^2(\mathbf{r}_\perp-\mathbf{r}_{0,\perp})}{L} e^{i(k_l-k_k)z}\hat{z} \label{eq:d_freest_relation},
\end{flalign}
where we have approximated the free electron states as a punctual distribution in a lateral direction and assumed that the electron's momentum doesn't change much within the interaction, and therefore the initial and final electron momentum are given approximately by the incident electron's central momentum, $k_0$. Of course this also means that we assume that the electron wavepacket has a well defined central momentum. Putting this into \Cref{eq:general_couplings_lightmatter}
\begin{flalign}
g_{kl}^{e-c}=\frac{e k_0}{mL}\sqrt{\frac{\hbar \mu_0}{\pi}\iint dz\, dz'\, \Im\Big\{\hat{z}\cdot\overline{\overline{\mathcal{G}}}([\mathbf{r}_{\perp,0},z],[\mathbf{r}_{\perp,0},z'],\omega_c) \cdot\hat{z}\Big\} e^{i(k_l-k_k)(z-z')} }
\end{flalign}
 For an electron interacting with an arbitrary system, initially in the ground state, the first order interaction term will give a probability for the electron to loose a certain amount of momentum proportional to $(g_{kl}^{e-c})^2$. It is then nice to check that the classical EELS probability\cite{GarciadeAbajo2010} indeed has the form $(g_{kl}^{e-c})^2$ if one makes the change $k_l-k_k\rightarrow \omega/v_0$, with $v_0$ being the initial speed of the electron. We will later see that this substitution is just an implication of energy conservation in the photon exchange between the optical modes and the electron. The fact that we have related the coupling strength to the classical EM Green's function through MQED allows to apply this strategy to arbitrary optical modes and we have the certainty that the normalization of the bosonic operators will be the right one by construction, which implies that this coupling strength expressions are physical.
 
 Also note that the coupling strength above only depends on the momentum of the electron through the momentum change over the interaction. By writing $k_k\equiv k$ and $q\equiv k_l-k_k$ one can see that the above coupling strength can be parametrized just as $g_{q}^{e-c}$. The same applies to the optical modes defined in \Cref{eq:currentcentered_a_single_freq}, albeit with the peculiarity that since the cavity's Green function is purely imaginary at the resonance frequency (See Section \ref{sec_sm:QS_Sphere_GF_derivation}), then one can show from \Cref{eq:currentcentered_a_single_freq} that $\hat{a}_q=-\hat{a}_{-q}$. With this, the interaction term between a free electron and an optical mode is given by 
 \begin{flalign}
    \hat{H}_I^{e-c}&=\hbar\sum_{q}g_{q}^{e-c} \hat{b}_q \left(\hat{a}_{-q}^\dagger-\hat{a}_{-q}\right) \nonumber\\
    &\approx \hbar\sum_{q } g_{q}^{e-c}  \hat{b}_q \left(\hat{a}_{-q_0}^\dagger-\hat{a}_{-q_0}\right) \mbox{sign}(q) \label{eq:MQED_lm_electron_doupling_gemeral_Hamiltonian}\\
    \hat{a}_{q}&=\frac{-i k_0 e\sum_\lambda}{2m\omega L g_{q}} \iint dz\, d\mathbf{r}'\, ^{i q z}\hat{z}\cdot \overline{\overline{\mathcal{G}}}_\lambda([\mathbf{r}_{\perp,0},z],\mathbf{r}',\omega_c)\cdot \hat{f}_\lambda(\mathbf{r}',\omega_c)\\
    g_{q}^{e-c}&=\frac{e k_0}{mL}\sqrt{\frac{\hbar \mu_0}{\pi}\iint dz\, dz'\, \Im\Big\{\hat{z}\cdot\overline{\overline{\mathcal{G}}}([\mathbf{r}_{\perp,0},z],[\mathbf{r}_{\perp,0},z'],\omega_c) \cdot\hat{z}\Big\} e^{iq(z-z')} } \label{eq:electron-cavity_final_coupling}
\end{flalign}
Where we have approximated that the optical modes that the electron will interact with are given by those that exchange momentum $q_0 =\omega_c/v_0$. Upon studying the interaction of an electron with an isolated optical mode and enforcing energy conservation, this assertion can be seen to be correct. When considering the interaction of an electron with a polaritonic system, the energy exchanges will differ from $\omega_c$, but since optical plamonic modes are highly spatially confined, their reciprocal space content will be very wide momenta distribution, and therefore, the error included by naming all the optical modes as $\hat{a}_{-q_0}$ is negligible.\\

In passing, note that the coupling strength above shows that a simple matching condition between the optical modes and the momentum exchange allows to maximize the coupling strength: for an electron interacting with an optical mode of frequency $\omega_c$, energy conservation imposes $q=\omega_c/v_0$. The coupling strength above involves a fourier transform of the associated field profiles, and therefore, for constructive interference one would want that the spatial frequency of the optical modes matches that of $\exp(i q z)$. If we consider a guided mode inside a waveguide, with guided momentum $k=k_0 n_{eff}$, where $n_{eff}$ is the effective refractive index of the guided mode, then we see that by making $k=q=\omega_c/v_0\rightarrow v_0 = c/n_{eff}$, will lead to enhanced interaction strength. This amounts to matching the phase velocity of the optical mode in the waveguide to that of the momentum exchange of the free electron, just as done in \cite{Kfir2020}.

\subsection{Bound electron - free electron interaction}
From \Cref{eq:effectiveHamiltonian_complete}, and having particularized the current densities for transitions in dipolar QE and free electrons, the hamiltonian describing the free-electron bound electron interaction is given by:
\begin{flalign*}  
\hat{H}_I^{e-QE}=-2\hbar\sum_{kl}\left[g_{ge,lk}^{e-QE}\,\hat{\sigma}+(g_{ge,kl}^{e-QE})^*\,\hat{\sigma}^\dagger\right]\sigma_{kl},
\end{flalign*}
where we have introduced the ladder operators of the QE given by $\hat{sigma}=\dyad{g}{e}$, and explicit expressions for both couplings in the last term are: 
\begin{gather*}
    g_{ge,lk}^{e-QE}=i k_0\frac{e^2 \hbar \mu_0}{2 m^2 L}\frac{m\omega_{ge}}{e\hbar}  \int d z'\,\boldsymbol{\mu}_{ge}\cdot \mbox{Re}\left\{\overline{\overline{G}}\left(\mathbf{r}_{QE},[\mathbf{r}_{0,\perp},z'],|\omega_{ge}|\right)\right\}\cdot \hat{z} e^{i(k_l-k_k)z'},\\
    g_{ge,kl}^{e-QE}=i k_0\frac{e^2 \hbar \mu_0}{2 m^2 L}\frac{m\omega_{ge}}{e\hbar}  \int d z'\,\boldsymbol{\mu}_{ge}\cdot \mbox{Re}\left\{\overline{\overline{G}}\left(\mathbf{r}_{QE},[\mathbf{r}_{0,\perp},z'],|\omega_{ge}|\right)\right\}\cdot \hat{z} e^{-i(k_l-k_k)z'},
\end{gather*}
Defining $k_k=k$ and $k_l=k_k+q$, we see that  the two couplings become independent of $k$
\begin{gather*}
    g_{q}^{e-QE}\equiv 2 g_{ge,lk}^{e-QE}=i k_0\frac{e^2 \hbar \mu_0}{m^2 L}\frac{m\omega_{ge}}{e\hbar}  \int d z'\,\boldsymbol{\mu}_{ge}\cdot \mbox{Re}\left\{\overline{\overline{G}}\left(\mathbf{r}_{QE},[\mathbf{r}_{0,\perp},z'],|\omega_{ge}|\right)\right\}\cdot \hat{z} e^{i q z'},\\
    2 g_{ge,kl}^{e-QE}\equiv g_{-q}^{e-QE}=-(g_{q}^{e-QE})^*,
\end{gather*}
which allows to parametrize the interaction hamiltonian entirely in terms of the momentum exchange between the free electron and the QE: $q$. With this, the interaction hamiltonian and coupling strengths read
\begin{flalign}  
\hat{H}_I^{e-QE}&=-\hbar\sum_{q}g_{q}^{e-QE}\left[\hat{\sigma}-\hat{\sigma}^\dagger\right] \hat{b}_q,\label{eq:H_int_e_QE_simplified}\\
g_{q}^{e-QE}&=i k_0\frac{e \mu_0 \omega_{QE}}{ m L} \int d z'\,\boldsymbol{\mu}\cdot \mbox{Re}\left\{\overline{\overline{G}}\left(\mathbf{r}_{QE},[\mathbf{r}_{0,\perp},z'],|\omega_{QE}|\right)\right\}\cdot \hat{z} e^{i q z'},\nonumber
\end{flalign}
Notice that the coupling between the QE and the passing electron depends on the real part of the greens function. In our study, the QE and cavity will be in resonance. At the resonance frequency the GF of the NP becomes purely imaginary, and therefore the coupling between the QE and the electron will be mediated by the vacuum GF \cite{Novotny2009}. In the near field, it reads:
\begin{gather*}
    \overline{\overline{G_0}}(\mathbf{r},\mathbf{r}')\overset{NF}{\approx} \frac{e^{ikR}}{4\pi R } \frac{1}{k^2 R^2}\left[  -\overline{\overline{I}} +3\frac{\mathbf{R}\mathbf{R}}{R^2}\right].
\end{gather*}
where $\mathbf{R}=\mathbf{r}-\mathbf{r}'$, $R=|\mathbf{R}|$ and $\mathbf{R}\mathbf{R}$ denotes the outer product. To calculate $\overline{\overline{G}}(\mathbf{r}_{QE},[\mathbf{r}_\perp,z],\Omega)$, we have $\mathbf{r}_{QE}=[x_{QE},0,z_{QE}]$ $\mathbf{r}_{e}=[x_{QE}+b_{e-QE},0,z]$, so $\mathbf{R}=\mathbf{r}_{QE}-\mathbf{r}_e=-[b_{e-QE},0,z-z_{QE}]$. If we define $\hat{u}=\mathbf{R}/R$ we then have
\begin{gather*}
\boldsymbol{\mu} \cdot\overline{\overline{G}}_0(\mathbf{r}_{QE},[\mathbf{r}_\perp,z])\cdot \hat{z}= \frac{e^{ik\sqrt{b_{e-QE}^2+(z-z_{QE})^2}}}{4\pi\sqrt{b_{e-QE}^2+(z-z_{QE})^2} } \frac{ -\boldsymbol{\mu}\cdot \hat{z} +3(\boldsymbol{\mu}\cdot\hat{u})(\hat{u}\cdot \hat{z})}{k^2 \left[b_{e-QE}^2+(z-z_{QE})^2\right]} \\
\overset{QS}{\approx} \frac{ -(\boldsymbol{\mu}\cdot \hat{z})}{4\pi k^2 \left[b_{e-QE}^2+(z-z_{QE})^2\right]^{3/2}}+
\frac{3(\boldsymbol{\mu}\cdot\hat{x}) b_{e-QE}(z-z_{QE})}{4\pi k^2 \left[b_{e-QE}^2+(z-z_{QE})^2\right]^{5/2}}
+\frac{3(\boldsymbol{\mu}\cdot\hat{z})(z-z_{QE})^2}{4\pi k^2 \left[b_{e-QE}^2+(z-z_{QE})^2\right]^{5/2}},
\end{gather*}
Where we have used the quasi-static (QS) limit. So the coupling between the free electron and the 2 level system is given by
\begin{gather*}
g_{q}^{e-QE}=i k_0\frac{e \mu_0 \omega_{QE}}{4\pi m L  k^2}   \frac{e^{i q z_{QE}}\boldsymbol{\mu}}{|b_{e-QE}|^2} \cdot \Big\{ \Big[ 2 I_2(q b_{e-QE})-I_0(q b_{e-QE}) \Big] \hat{z}+\Big[3\,\mbox{sign}(b_{e-QE} q)I_1(q b_{e-QE})\Big]\hat{x} \Big\},
\end{gather*}
where we have defined the integrals
\begin{gather}
    I_n(\phi)=\int_{-\infty }^{\infty} dz \frac{z^n}{\left[1+z^2 \right]^{5/2}}e^{i|\phi|z} \label{eq:Useful_integrals_GF_couplings}
\end{gather}
which evaluate to 
\begin{flalign}
I_0(\phi)=&\frac{2}{3} \left\lvert \phi\right\lvert^2 K_2(\left\lvert \phi\right\lvert),\\
I_1(\phi)=&\frac{2i}{3} \left\lvert \phi\right\lvert^2 K_1(\left\lvert \phi\right\lvert),\\
I_2(\phi)=&\frac{2}{3}\left[ \left\lvert\phi\right\lvert K_1(\left\lvert\phi\right\lvert)-\left\lvert\phi\right\lvert^2 K_0(\left\lvert\phi\right\lvert) \right],\\
I_0(\phi)-2I_2(\phi)=&2\left\lvert\phi\right\lvert^2 K_0(\left\lvert\phi\right\lvert). \nonumber
\end{flalign}
where in the last line we have used the recurrence relation \cite{Abramowitz1988} of modified bessel functions to rewrite the expression in a more compact form. From this, the interaction strength between the bound and free electron can be written as:
\begin{gather}
g_{q}^{e-QE}= -\frac{e\, k_0 \left\lvert q \right\lvert^2 e^{i q z_{QE}}}{2 \pi m L  \epsilon_0\omega_{QE}}    \boldsymbol{\mu} \cdot
\begin{pmatrix}
    \mbox{sign}(q b_{e-QE}) K_1(\left\lvert q b_{e-QE}\right\lvert)\\
    0\\
    i K_0(\left\lvert q b_{e-QE}\right\lvert)
    \end{pmatrix}. \label{eq:coupling_free_electron-bound_electron}
\end{gather}
Comparing this to a previous result \cite{Zhao2020} studying the interaction of free and bound electrons, we see that the expression agree, while our expression puts emphasis on the fact that the phase accumulated in the electron interaction with the QE will be different for a situation of energy loss and energy gain. 

\section{Quasi-static Greens Function of a Sphere}
\label{sec_sm:QS_Sphere_GF_derivation}
Here, we aim at obtaining a closed expression for the $\mathcal{G}(\mathbf{r},\mathbf{r}',\omega_c)$ tensor in the Green’s function decomposition for the optical mode supported by a spherical cavity:
\begin{gather*}
    \mathbf{G}(\mathbf{r},\mathbf{r'},\omega)=\mathbf{G}_0(\mathbf{r},\mathbf{r'},\omega)+\mathbf{\mathcal{G}}(\mathbf{r},\mathbf{r'})\mathcal{L}(\omega)
\end{gather*}
where $\mathbf{G}_0(\mathbf{r},\mathbf{r'},\omega)$ is the vacuum green's function and $\mathcal{L}(\omega)$ contains the frequency response of the sphere's contribution to the total GF.
\begin{figure}[]
\centering
\includegraphics[angle=0,width=0.55\columnwidth]{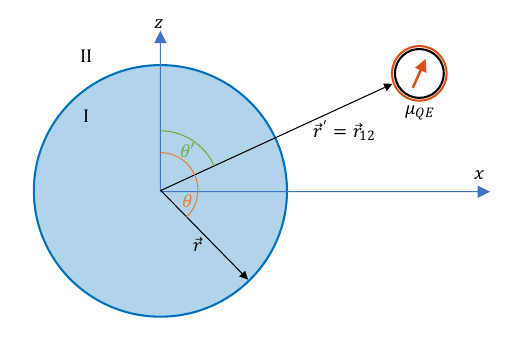}
\caption{Sketch of the system under consideration.} \label{Appendixfig:fig1}
\end{figure}
To do this we solve Poisson equation in the geometry presented in \Cref{Appendixfig:fig1} with the charge distribution of an electric dipole:
\begin{gather*}
    \rho(\mathbf{r}',\omega)=-\mathbf{\mu}\cdot \mathbf{\nabla}\mathbf{\delta}(\mathbf{r}'-\mathbf{r}_\mu)
\end{gather*}
We use that for $r'>r$, we can expand
\begin{gather*}
    \frac{1}{|\mathbf{r}-\mathbf{r}'|}=\sum_{n=0}^{\infty}\sum_{m=-n}^{n}\frac{4\pi}{2n+1}{Y_n^m}^*(\theta',\phi')Y_n^m(\theta,\phi)\frac{r^n}{{r'}^{n+1}}
\end{gather*}
Which allows to express the potential created by the dipole as
\begin{flalign*}
    \phi_{inc}(\mathbf{r},\omega)=\frac{1}{4\pi\epsilon_0}\int  d\mathbf{r}'  \frac{\rho(\mathbf{r}',\omega)}{|\mathbf{r}-\mathbf{r}'|}=\frac{1}{4\pi\epsilon_0}\sum_{n=0}^{\infty}\sum_{m=-n}^{n}\frac{4\pi}{2n+1}I_n^m Y_n^m(\theta,\phi)r^n
\end{flalign*}
with 
\begin{gather*}
    I_n^m=\int d\mathbf{r}' \frac{\rho(\mathbf{r}',\omega)}{{r'}^{n+1}}{Y_n^m}^*(\theta',\phi')=-\mathbf{\nabla}'\left( \frac{{Y_n^m}^*(\theta',\phi')}{{r'}^{n+1}}\right)\cdot \boldsymbol{\mu}
\end{gather*}
Now we can solve the scattering problem by expanding the potential on the different regions and matching spherical harmonics:
\begin{flalign*}
{\phi_I}_n^m(\mathbf{r},\omega)=&{\phi_{inc}}_n^m(\mathbf{r},\omega)+{\phi_{sc,I}}_n^m(\mathbf{r},\omega)
=\frac{1}{\epsilon_0}\frac{1}{2n+1}I_n^m Y_n^m(\theta,\phi)r^n + A_n^m \frac{Y_n^m(\theta,\phi)}{r^{n+1}}\\
{\phi_{II}}_n^m(\mathbf{r},\omega)=&{\phi_{sc,II}}_n^m(\mathbf{r},\omega)=B_n^m Y_n^m(\theta,\phi)r^{n}
\end{flalign*}
Imposing continuity of the potential and the normal component of the displacement field we find $A_n^m$, and hence the scattered field
\begin{gather*}
    {\phi_{sc,I}}_n^m(\mathbf{r},\omega)=-\left[\frac{1}{\epsilon_0}\frac{R^{2n+1}}{2n+1} I_n^m \frac{\epsilon_{sph}(\omega)-1}{\epsilon_{sph}(\omega)+\frac{n+1}{n}}\right]\frac{Y_n^m(\theta,\phi)}{r^{n+1}}
\end{gather*}
Note that all the frequency dependence is contained within the permittivity of the sphere. If we assume a Drude model for the sphere $\epsilon_{sph}=1-\omega_p^2/(\omega(\omega+i\gamma))$, then we can rewrite the permittivity quotient as
\begin{gather*}
    \frac{\epsilon_{sph}(\omega)-1}{\epsilon_{sph}(\omega)+\frac{n+1}{n}}=\frac{-\frac{\omega_p^2}{\omega(\omega+i\gamma)}}{\frac{2n+1}{n}-\frac{\omega_p^2}{\omega(\omega+i\gamma)}}=\frac{-\omega_{sp}^2}{\omega(\omega+i\gamma)-\omega_{sp}^2}
\end{gather*}
where we have defined $\omega_c=\omega_{sp}=\sqrt{n/(2n+1)}\,\omega_p$, the generalized Frölich poles. The above quotient has poles for $\omega=\pm\sqrt{\omega_{sp}^2-(\gamma/2)^2}-i\gamma/2$, which  in the limit of small losses tends to $\omega_{sp}$. This means that within the good resonator approximation, the above quotient will be only non-zero for a small region around $\omega_{sp}$. Expressing $\omega = \omega_{sp}+\delta$, and expanding for small $\delta$ values we find:
\begin{gather*}
    \frac{\epsilon_{sph}(\omega)-1}{\epsilon_{sph}(\omega)+\frac{n+1}{n}}=\frac{-\omega_{sp}^2}{(\omega_{sp}+\delta)^2 +i\gamma(\omega_{sp}+\delta)-\omega_{sp}^2}\approx-\frac{1}{2}\frac{\omega_{sp}}{\omega-\omega_{sp}+i\frac{\gamma}{2}}=\frac{-\omega_{sp}}{2}\mathcal{L}(\omega,\omega_{sp},\gamma)
\end{gather*}
Which gives a lorentzian lineshape around every frölich pole, with $\mathcal{L}(\omega,\omega_{sp},\gamma)=(\omega-\omega_{sp} +i\gamma/2)^{-1}$. The scattered potential and corresponding field are then
 
\begin{flalign*}
    {\phi_{sc,I}}_n^m(\mathbf{r},\omega)&=\frac{1}{2\epsilon_0} \frac{R^{2n+1}}{2n+1}\, \omega_{sp}\mathcal{L}(\omega,\omega_{sp},\gamma)\, \frac{Y_n^m(\theta,\phi)}{r^{n+1}} \mathbf{\nabla}'\left( \frac{{Y_n^m}^*(\theta',\phi')}{r'^{n+1}}\right)\cdot \boldsymbol{\mu}
    \\
    {\mathbf{E}_{sc,I}}_n^m(\mathbf{r},\omega)&=-\frac{1}{2\epsilon_0} \frac{R^{2n+1}}{2n+1}\, \omega_{sp}\mathcal{L}(\omega,\omega_{sp},\gamma)\, 
    \mathbf{\nabla}\left(\frac{Y_n^m(\theta,\phi)}{r^{n+1}}\right)\mathbf{\nabla}'\left( \frac{{Y_n^m}^*(\theta',\phi')}{r'^{n+1}}\right)\cdot \boldsymbol{\mu}
\end{flalign*}
 
By definition the Dyadic Green's function fulfil
\begin{gather*}
    \mathbf{E}_\mu=\frac{\omega^2}{c^2\epsilon_0}\mathbf{G}(\mathbf{r},\mathbf{r}',\omega))\cdot \boldsymbol{\mu}
\end{gather*}
and therefore the contributions of all the different spherical harmonics to the greens functions are:
\begin{flalign*}
    \mathbf{G}_n^m(\mathbf{r},\mathbf{r}',\omega)=-\frac{c^2}{\omega^2}\frac{R^{2n+1}}{2(2n+1)}\, \omega_{sp}\mathcal{L}(\omega,\omega_{sp},\gamma)\mathbf{\nabla}\left(\frac{Y_n^m(\theta,\phi)}{r^{n+1}}\right) \otimes \mathbf{\nabla}'\left( \frac{{Y_n^m}^*(\theta',\phi')}{r'^{n+1}}\right)
\end{flalign*}
In particular we are interested in the contributions given by the dipolar mode of the nanoparticle, corresponding to $l=1$. Setting $l=1$ inmediately gives three possible values for $m$, which gives the expected three-fold degeneracy of the dipolar mode. Instead of using the complex formulation, we choose to work with the real-valued spherical harmonics, given by
\begin{flalign*}
\mathcal{Y}_x&=\frac{1}{\sqrt{2}}\left[Y_1^{-1}-Y_1^{1}\right]=\frac{1}{2}\sqrt{\frac{\pi}{3}}\frac{x}{r}\\
\mathcal{Y}_y&=\frac{1i}{\sqrt{2}}\left[Y_1^{-1}+Y_1^{1}\right]=\frac{1}{2}\sqrt{\frac{\pi}{3}}\frac{y}{r}\\
\mathcal{Y}_z&=Y_1^{0}=\frac{1}{2}\sqrt{\frac{\pi}{3}}\frac{z}{r}
\end{flalign*}
So the greens functions can be expressed as
\begin{gather*}
  \mathbf{G}_{x,y,z}(\mathbf{r},\mathbf{r}',\omega)=-\frac{c^2}{\omega^2}\frac{\pi R^3}{72}\, \omega_{sp}\mathcal{L}(\omega,\omega_{sp},\gamma)
     \mathbf{\nabla}\left(\frac{x,y,z}{r^3}\right) \otimes \mathbf{\nabla}'\left( \frac{x',y',z'}{r'^3}\right)
\end{gather*}
To express this dyadic in the form of \Cref{eq:resonant_GF_aux}, it is enough to note that in the limit of vanishing absorption $\lim_{\gamma\rightarrow0}\mathcal{L}(\omega,\omega_{sp},\gamma)= -i\pi \delta(\omega-\omega_{sp})$, and that $\mathcal{L}(\omega_{sp},\omega_{sp},\gamma)=-2i/\gamma$. 
 
\begin{flalign}
\mathbf{G}_{x,y,z}(\mathbf{r},\mathbf{r}',\omega)&= \mathbf{\mathcal{G}}_{x,y,z}(\mathbf{r},\mathbf{r}',\omega_{sp}) \delta(\omega-\omega_{sp}),\\
\mathbf{\mathcal{G}}_{x,y,z}(\mathbf{r},\mathbf{r}',\omega_{sp})&= -\frac{\gamma\pi}{2} \frac{c^2}{\omega_{sp}^2}\frac{\pi R^3}{72}\, \omega_{sp} \mathcal{L}(\omega_{sp},\omega_{sp},\gamma) 
     \mathbf{\nabla}\left(\frac{x,y,z}{r^3}\right) \otimes \mathbf{\nabla}'\left( \frac{x',y',z'}{r'^3}\right).
\end{flalign}
Finally, noting that $\partial_i(x_j/r^3)=(r^2\delta_{i,j}-3x_ix_j)/r^5$, the entries of the Greens function of the three degenerated dipolar cavity modes (denoted by $x_k$) can be written as:
\begin{flalign}
[\mathbf{\mathcal{G}}_{x_k}]_{ij}(\mathbf{r},\mathbf{r}',\omega_{sp})&= i\pi \frac{c^2}{\omega_{sp}}\frac{\pi R^3}{72}\,
     \partial_i\left(\frac{x_k}{r^3}\right) \otimes \partial_j'\left( \frac{x_k'}{r'^3}\right)\nonumber\\
     &= i\pi \frac{c^2}{\omega_{sp}}\frac{\pi R^3}{72}\, \frac{(r^2\delta_{i,k}-3x_ix_k)({r'}^2\delta_{j,k}-3x_j'x_k')}{r^5 {r'}^5},
\end{flalign}

\subsection{Evaluation of QE - cavity coupling}
By using \Cref{eq:MQED_lm_dip_doupling_gemeral}, and the Green's function of a sphere from above, the coupling between a QE of dipole moment $\boldsymbol{\mu}$, when the dipole is placed on the $\hat{x}$ axis at a distance $b_{c-QE}$ from the nanoparticle is given by: 
\begin{flalign} 
g^{c-QE}_x&=\frac{\left\lvert \omega_{QE}\right\rvert}{3}\left\lvert \boldsymbol{\mu}\cdot \hat{x}\right\lvert \sqrt{ \frac{\pi}{2}\left( \frac{R}{b_{c-QE}}\right)^3 \frac{ 1}{\hbar\omega_{sp}} \,\frac{1}{\epsilon_0\,b_{c-QE}^3}} \label{eq:MQED_dip_dip_cavity_coupl_x}\\
g^{c-QE}_y&=\frac{\left\lvert \omega_{QE}\right\rvert}{6}\left\lvert \boldsymbol{\mu}\cdot \hat{y}\right\lvert\sqrt{ \frac{\pi}{2}\left( \frac{R}{b_{c-QE}}\right)^3 \frac{ 1}{\hbar\omega_{sp}} \, \frac{1}{\epsilon_0\,b_{c-QE}^3}} \label{eq:MQED_dip_dip_cavity_coupl_y}\\
g^{c-QE}_z&=\frac{\left\lvert \omega_{QE}\right\rvert}{6}\left\lvert \boldsymbol{\mu}\cdot \hat{z}\right\lvert\sqrt{ \frac{\pi}{2}\left( \frac{R}{b_{c-QE}}\right)^3 \frac{ 1}{\hbar\omega_{sp}} \, \frac{1}{\epsilon_0\,b_{c-QE}^3}} \label{eq:MQED_dip_dip_cavity_coupl_z}
\end{flalign}
which shows that for a dipole oriented along one of the coordinate axis, it only couples to one of the three degenerated dipolar modes of the cavity.

\subsection{Evaluation of free electron - cavity coupling}
In the same way, using \Cref{eq:electron-cavity_final_coupling} and the Green's function of a spherical nanoparticle we find that an electron that passes a distance $\mathbf{r}_{\perp,0}=b_{e-c}\hat{x}$ of the cavity has a coupling to the different dipolar modes given by: 
\begin{flalign*}
    g_{q,x}^{e-c}&=\frac{e k_0}{mL}\sqrt{\frac{\hbar}{\epsilon_0 \omega_{sp}}\frac{\pi R^3}{8} \frac{1}{|b_{e-c}|^4} I_1(q b_{e-c})I_1^*(q b_{e-c})}\\
    g_{q,y}^{e-c}&=0\\
    g_{q,z}^{e-c}&=\frac{e k_0}{3mL}\sqrt{\frac{\hbar}{\epsilon_0 \omega_{sp}}\frac{\pi R^3}{8} \frac{1}{|b_{e-c}|^4}\left[I_0(q b_{e-c})-2I_2(q b_{e-c})\right]\left[I_0(q b_{e-c})-2I_2(q b_{e-c})\right]^* }
\end{flalign*}
where we have used the integrals defined in \Cref{eq:Useful_integrals_GF_couplings}. From this, the interaction strength between the passing electron and the dipolar modes of the cavity is given by
\begin{flalign}
    g_{q,x}^{e-c}&=\frac{e\hbar  k_0}{3mL} \left\lvert q\right\lvert^2  K_1(\left\lvert q b_{e-c} \right\lvert) \sqrt{\frac{1}{\epsilon_0 \hbar\omega_{sp}}\frac{\pi R^3}{2}  } \label{eq:MQED_electron_dip_cavity_coupl_x}\\
    g_{q,y}^{e-c}&=0 \label{eq:MQED_electron_dip_cavity_coupl_yx}\\
    g_{q,z}^{e-c}&=\frac{e\hbar k_0}{3mL} \left\lvert q\right\lvert^2 K_0(\left\lvert q b_{e-c}\right\lvert)\sqrt{\frac{1}{\epsilon_0 \hbar \omega_{sp}}\frac{\pi R^3}{2}   } \label{eq:MQED_electron_dip_cavity_coupl_z}
\end{flalign}
Having seen before the coupling strength between a free electron and a QE (\Cref{eq:coupling_free_electron-bound_electron}), we can extract the dipole moment that the free electron induces in the cavity:
\begin{gather}
 \left\lvert \boldsymbol{\mu}_{c} \right\lvert =\frac{2\pi}{3} \sqrt{\pi R^3\,\hbar \omega_{sp} \epsilon_0}
\end{gather}
We further note that the amplitudes of this dipole moment satisfy: $|\boldsymbol{\mu}_c\cdot\hat{x}|=|\boldsymbol{\mu}_c\cdot\hat{z}|$ and $\boldsymbol{\mu}_c\cdot\hat{y}=0$. From here one can see that for the parameters used in the main text $|\boldsymbol{\mu}_{c}^{ind}|\approx 40 |\boldsymbol{\mu}_{QE}|$. This expression agrees well with the one derived from comparing the polarizability of a sphere with that of a quantum 2-level system \cite{Ballestero2015}.\\

If one calculates the integrated loss probability of an electron interacting with a spherical metallic nanoparticle in the non-retarded limit \cite{GarciadeAbajo2010}, and applies the good resonator approximation to the drude description, the loss probability is calculated as:
\begin{gather}
    P_L=\int d\omega \Gamma_{NR}^{sph}(\omega)=\int d\omega \frac{ e^2}{\pi^2 \epsilon_0 \hbar v_0^2} \left[ \left( \frac{\omega}{v}\right)^2 K_0^2\left( \frac{\omega b}{v}\right)+\left( \frac{\omega}{v}\right)^2 K_1^2\left( \frac{\omega b}{v}\right)\right] \mbox{Im}\left\{\alpha_{NR} (\omega)\right\}\\
    \alpha_{NR} (\omega)=R^3\frac{\epsilon(\omega)-1}{\epsilon(\omega)+2}\approx i \frac{\omega_{sp}\pi R^3}{2}\delta(\omega-\omega_{sp})\\
     P_L= \frac{e^2}{\pi^2 \epsilon_0 \hbar \omega_{sp}} \left( \frac{\omega_{sp}^4}{v^4}\right)\left[  K_0^2\left( \frac{\omega_{sp} b}{v_0}\right)+ K_1^2\left( \frac{\omega_{sp} b}{v_0}\right)\right] \frac{\pi R^3}{2}
\end{gather}
Looking at the electron-cavity couplings above, and taking the limit of small coupling, one can see that the first order of interaction between a free electron and a spherical nanoparticle, with the nanoparticle initially in it's ground state yields a probability for the electron to loose one photon that looks like
\begin{flalign}
    P_L\approx& \left \lvert \frac{L}{v_0} g_{\omega_{sp}/v_0,x}^{e-c}\right\lvert^2 +\left \lvert \frac{L}{v_0} g_{\omega_{sp}/v_0,z}^{e-c}\right\lvert^2 \\
    =& \frac{e^2}{9 \epsilon_0 \hbar\omega_{sp}} \left\lvert \frac{\omega_{sp}}{v_0}\right\lvert^4  \left[ K_0^2\left( \left\lvert \frac{\omega_{sp} b_{e-c}}{v_0} \right\lvert\right)+K_1^2\left(\left\lvert \frac{\omega_{sp} b_{e-c}}{v_0} \right\lvert\right) \right] \frac{\pi R^3}{2}  
\end{flalign}
Which agrees with the classical result to a factor of $(\pi/3)^2$. The agreement between our formalism and the classical result is not a coincidence, and besides the analytical similarity of the calculations involved in the derivation, what matters is that through MQED we have been able to define properly normalized bosonic modes, which means that once the electromagnetic Green's function of an arbitrary system is known, the interaction can be studied in the way that we have outlined.


\section{Algebraic structure of scattering matrix and general properties of electrons interacting with quantum systems}
In the main text we show that the scattering matrix can be written as 
\begin{flalign}
     \hat{S}=&\exp\left(-i \hat{O} \right)\\
     \hat{O}=&\frac{L}{\hbar v_0}\sum_{\phi,\phi'}\delta_{E_{\phi},E_{\phi'}} \dyad{\phi} \hat{H}_{I}\dyad{\phi'}
\end{flalign}
with the interaction hamiltonian given by
\begin{flalign}
\hat{H}_I=&\hbar \sum_{i=x,z} \sum_{q} \, g_{q,i}^{c-e}\hat{b}_q \left[\hat{a}_i^\dagger-\hat{a}_i\right] \mbox{sign}(q)-\hbar\sum_{q}g_{q}^{e-QE}\hat{b}_q\left\{ \sigma - \sigma^{\dagger}\right\}=\\
    \equiv &\sum_q \hat{H}_{I,q} \hat{b}_q
\end{flalign}
where we have grouped all the operators that act on the cavity-QE system inside $\hat{H}_{I,q}$. We could then explicitly write out the $\ket{\phi}$ initial and final states in terms of the bare cavity ($\ket{n_z}$), polaritons ($\ket{N,\pm}$) and electronic momenta ($\ket{k}$) as $\ket{n_z, N,\pm,k}$, but for compactness, we choose to only explicitly write the electron state, and introduce $\ket{\psi}$ as a compact representation of the rest of the system. Then the $\hat{O}$ operator can be writen as: 
\begin{flalign}
    \hat{O}&=\frac{L}{\hbar v_0}\sum_{\begin{array}{c}
     \psi,k\\
     \psi',k'
  \end{array}}\delta_{E(\psi,k),E(\psi',k'))} \dyad{\psi,k} \hat{H}_{I}\dyad{\psi',k'}\\
  &=\frac{L}{\hbar v_0}\sum_{\begin{array}{c}
     \psi,k\\
     \psi',k'
  \end{array}}\sum_q \delta_{E(\psi,k),E(\psi',k'))} \ket{\psi,k} \bra{\psi,k} \hat{H}_{I,q} \ket{\psi',k'-q}\bra{\psi',k'}\\
  &=\frac{L}{\hbar v_0}\sum_{\begin{array}{c}
     \psi\\
     \psi',k'
  \end{array}}\sum_q \delta_{E(\psi,k'-q),E(\psi',k'))} \dyad{\psi} \hat{H}_{I,q} \dyad{\psi'}\otimes \ket{k'-q}\bra{k'}
\end{flalign}
Using the free electron energy dispersion, and the non-recoil approximation, one can see that energy conservation imposes 
\begin{gather}
    q\equiv q_{\psi,\psi'}=\frac{E(\psi)-E(\psi')}{\hbar v_0}
\end{gather}
with $v_0$ being the central speed of the incident electron. This is independent of the initial momentum $k'$, and therefore the evolution operator $\hat{O}$ can be cast as: 
\begin{gather}
    \hat{O}=\frac{L}{\hbar v_0}\sum_{\psi,\psi'} \dyad{\psi} \hat{H}_{I,q_{\psi,\psi'}}\otimes \hat{b}_{q_{\psi,\psi'}} \dyad{\psi'}\equiv \sum_{\psi,\psi'} \mathcal{h}_{I,\psi,\psi'} \dyad{\psi}{\psi'}\otimes \hat{b}_{q_{\psi,\psi'}}\\
    \mathcal{h}_{I,\psi,\psi'}=\frac{L}{\hbar v_0}\bra{\psi}\hat{H}_{I,q_{\psi,\psi'}}\ket{\psi'}
\end{gather}
Which shows that through energy conservation, each possible transition within the reduced system is accompanied by a corresponding shift in the electron. This comes from the particular algebra of the electron's ladder operators and the non-recoil approximation alone, and therefore the expression above applies to electrons interacting with any particular kind of quantum system. Since the scattering matrix describing the final state of the system after the interaction of the electron with a QS is given by the exponentiation of the $\hat{O}$ operator, and we expect small interaction strength between electron-QS, it makes sense to take a look at the behavior of powers of $\hat{O}$. If we denote the basis of eigenstates of the quantum system by $\ket{\psi_i}$, then some powers can be written as:
\begin{flalign}
    \hat{O}&=\sum_{\psi_i,\psi_j} \mathcal{h}_{I,\psi_i,\psi_j} \dyad{\psi_i}{\psi_j}\otimes \hat{b}_{q_{\psi_i,\psi_j}}\\
    \hat{O}^2&=\sum_{\psi_i,\psi_j} \left[\sum_{\psi_k}
    \mathcal{h}_{I,\psi_i,\psi_k}\mathcal{h}_{I,\psi_k,\psi_j} \right]\dyad{\psi_i}{\psi_j}\otimes \hat{b}_{q_{\psi_i,\psi_j}}\\
    \hat{O}^3&=\sum_{\psi_i,\psi_j}\left[ \sum_{\psi_k,\psi_l} \mathcal{h}_{I,\psi_i,\psi_k}\mathcal{h}_{I,\psi_k,\psi_l}\mathcal{h}_{I,\psi_l,\psi_j} \right] \dyad{\psi_i}{\psi_j}\otimes \hat{b}_{q_{\psi_i,\psi_j}}\\
    \hat{O}^4&=\sum_{\psi_i,\psi_j} \left[ \sum_{\psi_k,\psi_l,\psi_m} \mathcal{h}_{I,\psi_i,\psi_k}\mathcal{h}_{I,\psi_k,\psi_l}\mathcal{h}_{I,\psi_l,\psi_m}\mathcal{h}_{I,\psi_m,\psi_j} \right] \dyad{\psi_i}{\psi_j}\otimes \hat{b}_{q_{\psi_i,\psi_j}}
\end{flalign}
Notation becomes cumbersome, but the main idea is that regardless of how many photon exchanges happen between the electron and QS (the power of operator $\hat{O}$ above), the commutativity of the electron ladder operators guarantee that the final shift that the electron experiences will only depend on the energy difference between the initial and final states connected by the interaction, and therefore $\hat{O}^n$ will always have the form, 
\begin{gather}
    \hat{O}^n =\sum_{\psi_i,\psi_j} \mathcal{h}_{I,\psi_i,\psi_j}^{(n)} \dyad{\psi_i}{\psi_j}\otimes \hat{b}_{q_{\psi_i,\psi_j}}
\end{gather}
And therefore, the general form of the propagator becomes: 
\begin{gather}
    \hat{{S}}=\exp\left(-\frac{i}{\hbar} \frac{L}{v} \hat{O} \right)=\sum_{\psi_i,\psi_j} \left[ \sum_{n=0}^\infty \left(-\frac{i}{\hbar} \frac{L}{v} \right)^n \frac{\mathcal{h}_{I,\psi_i,\psi_j}^{(n)} }{n!}  \right] \dyad{\psi_i}{\psi_j}\otimes \hat{b}_{q_{\psi_i,\psi_j}}\equiv\\
    \equiv  \sum_{\psi_i,\psi_j} \mathcal{S}_{\psi_i,\psi_j} \dyad{\psi_i}{\psi_j}\otimes \hat{b}_{q_{\psi_i,\psi_j}}
    \label{eqs:Scatt_matt_general}
\end{gather}
This structure has direct implications on how the interaction with the electron might create or modify populations on any QS. Suppose for instance that initially the QS is prepared on one of the basis states, $\ket{\psi_n}$, and the electron is prepared with an arbitrary momentum distribution $\ket{\psi_e}=\int dk B(k) \ket{k}$. The final state of the system after interaction will be given by $\ket{\psi}=\sum_{\psi_i} S_{\psi_i,\psi_n} \hat{b}_{q_{\psi_i,\psi_n}}\ket{\psi_i,\psi_e}$, and the corresponding density matrix, and reduced density matrix for the QS are given by: 
\begin{gather}
    \rho=\dyad{\psi}=\sum_{\psi_i,\psi_j} \mathcal{S}_{\psi_i,\psi_n} \mathcal{S}_{\psi_j,\psi_n}^* \hat{b}_{q_{\psi_i,\psi_n}} \dyad{\psi_i,\psi_e}{\psi_j,\psi_e}\hat{b}_{q_{\psi_j,\psi_n}}^\dagger\\
     \rho^{QS}=\sum_{\psi_i,\psi_j} \mathcal{S}_{\psi_i,\psi_n} \mathcal{S}_{\psi_j,\psi_n}^* \dyad{\psi_i}{\psi_j}\  \int dk B(k)B^*(k-\left[ q_{\psi_i,\psi_n}-q_{\psi_j,\psi_n}\right])
\end{gather}
The second expression above, despite the absolute agnosticity about the particular QS the electron interacts with, allows to assert that if the QS is initially in only one of it's eigenstates, then the final populations are absolutely independent of the particular electronic wavefunction and only the coherences will be sensitive to the initially prepared electron. Now instead consider that the QS is initially prepared in a superposition of two of it's basis states: $\ket{\psi_{n_1}}$ and $\ket{\psi_{n_2}}$ as $\ket{\psi^{QS}_0}=\cos(\theta)\ket{\psi_{n_1}}+\sin(\theta)e^{i\phi}\ket{\psi_{n_2}}$. Then the final state $\ket{\psi}$, density matrix $\rho$ and reduced density matrix $\rho^{QS}$ read:
\begin{gather}
    \ket{\psi}=\sum_{\psi_i}\left[ \cos(\theta)\mathcal{S}_{\psi_i,\psi_{n_1}}\hat{b}_{q_{\psi_i,\psi_{n_1}}}+\sin(\theta)e^{i\phi}\mathcal{S}_{\psi_i,\psi_{n_2}} \hat{b}_{q_{\psi_i,\psi_{n_2}}}\right]\ket{\psi_{i},\psi_e}
\end{gather}
\begin{flalign}
    \rho=\sum_{\psi_i,\psi_j}\Big\{ & \cos^2(\theta)\mathcal{S}_{\psi_i,\psi_{n_1}}\mathcal{S}^*_{\psi_j,\psi_{n_1}}\hat{b}_{q_{\psi_i,\psi_{n_1}}} \dyad{\psi_{i},\psi_e}{\psi_{j},\psi_e}\hat{b}_{q_{\psi_j,\psi_{n_1}}}^\dagger \nonumber\\
    +&\sin^2(\theta)\mathcal{S}_{\psi_i,\psi_{n_2}} \mathcal{S}^*_{\psi_j,\psi_{n_2}} \hat{b}_{q_{\psi_i,\psi_{n_2}}}\dyad{\psi_{i},\psi_e}{\psi_{j},\psi_e}\hat{b}_{q_{\psi_j,\psi_{n_2}}}^\dagger \nonumber\\
    +& \cos(\theta)\sin(\theta)e^{-i\phi}\mathcal{S}_{\psi_i,\psi_{n_1}}\mathcal{S}^*_{\psi_j,\psi_{n_2}}\,\hat{b}_{q_{\psi_i,\psi_{n_1}}}\dyad{\psi_{i},\psi_e}{\psi_{j},\psi_e}
    \hat{b}_{q_{\psi_j,\psi_{n_2}}}^\dagger\nonumber \\
    +&\cos(\theta)\sin(\theta)e^{i\phi}\mathcal{S}_{\psi_i,\psi_{n_2}}\mathcal{S}^*_{\psi_j,\psi_{n_1}}\,\hat{b}_{q_{\psi_i,\psi_{n_2}}}\dyad{\psi_{i},\psi_e}{\psi_{j},\psi_e}
    \hat{b}_{q_{\psi_j,\psi_{n_1}}}^\dagger\Big\}
\end{flalign}
\begin{flalign}
    \rho^{QS}=\sum_{\psi_i,\psi_j}\Big\{ & \cos^2(\theta)\mathcal{S}_{\psi_i,\psi_{n_1}}\mathcal{S}^*_{\psi_j,\psi_{n_1}} \int dk B(k) B^*(k-[q_{\psi_i,\psi_{n_1}}-q_{\psi_j,\psi_{n_1}}]) \nonumber\\
    +&\sin^2(\theta)\mathcal{S}_{\psi_i,\psi_{n_2}} \mathcal{S}^*_{\psi_j,\psi_{n_2}} \int dk B(k) B^*(k-[q_{\psi_i,\psi_{n_2}}-q_{\psi_j,\psi_{n_2}}]) \nonumber\\
    +& \cos(\theta)\sin(\theta)e^{-i\phi}\mathcal{S}_{\psi_i,\psi_{n_1}}\mathcal{S}^*_{\psi_j,\psi_{n_2}}\, \int dk B(k) B^*(k-[q_{\psi_i,\psi_{n_1}}-q_{\psi_j,\psi_{n_2}}])\nonumber \\
    +&\cos(\theta)\sin(\theta)e^{i\phi}\mathcal{S}_{\psi_i,\psi_{n_2}}\mathcal{S}^*_{\psi_j,\psi_{n_1}}\,\int dk B(k) B^*(k-[q_{\psi_i,\psi_{n_2}}-q_{\psi_j,\psi_{n_1}}]) \Big\} \dyad{\psi_i}{\psi_j}
\end{flalign}
Particularly, the populations are given by:
\begin{flalign}
    \bra{\psi_i}\rho^{QS}\ket{\psi_i}= & \cos^2(\theta)\mathcal{S}_{\psi_i,\psi_{n_1}}\mathcal{S}^*_{\psi_i,\psi_{n_1}} +\sin^2(\theta)\mathcal{S}_{\psi_i,\psi_{n_2}} \mathcal{S}^*_{\psi_i,\psi_{n_2}}\nonumber\\
    +& \cos(\theta)\sin(\theta)e^{-i\phi}\mathcal{S}_{\psi_i,\psi_{n_1}}\mathcal{S}^*_{\psi_i,\psi_{n_2}}\, \int dk B(k) B^*(k-q_{\psi_{n_2},\psi_{n_1}})\nonumber \\
    +&\cos(\theta)\sin(\theta)e^{i\phi}\mathcal{S}_{\psi_i,\psi_{n_2}}\mathcal{S}^*_{\psi_i,\psi_{n_1}}\,\int dk  B^*(k) B(k-q_{\psi_{n_2},\psi_{n_1}}) \\
    = & \cos^2(\theta)\left|\mathcal{S}_{\psi_i,\psi_{n_1}}\right|^2 +\sin^2(\theta) \left|\mathcal{S}_{\psi_i,\psi_{n_2}} \right|^2\nonumber\\
    +& \sin(2\theta)\, \mbox{Re}\left\{e^{-i\phi}\mathcal{S}_{\psi_i,\psi_{n_1}}\mathcal{S}^*_{\psi_i,\psi_{n_2}}\, \int dk B(k) B^*(k-q_{\psi_{n_2},\psi_{n_1}}) \right\} 
\end{flalign}
which explicitly shows that for arbitrary quantum systems, the electronic wavefunctions will give contributions to the final populations of the reduced system so long as the initial QS is found in a superposition of states, and the modulation of the electron is tuned appropriately to the energy difference between populated states.

\section{Computational implementation of the calculations}
The main advantage of the present work as compared to recent contributions to the study of polaritons through the use of electrons is that we have access to analytical expressions of the different quantities. Their particular expressions are too large to result useful to the general reader, but the way in which the calculation is implemented does have it's advantages. The main idea is to exploit two basic properties of the system: 
\begin{itemize}
    \item The initial state of the system can always be characterized as a product state of the electron and the rest of the system.
    \item The algebraic properties of the electronic ladder operators allows to treat them in a particularly useful way. 
\end{itemize}

Since the initial state of a free electron interacting with a quantum system (QS) can always be described as a product state between the electron and the QS, one can perform all the algebraic manipulations on the QS's Hilbert space and then apply the resulting $\hat{b}_q$ operators over the electronic wavefunction. In systems with only one energy scale, such as isolated quantum emitters or single mode fields, this strategy doesn't really save much work since the free electron's state can be characterized by just how many photons it has exchanged with the QS. However, when dealing with QS with several energy scales, then the electron can exchange photons of several different energies (which also depend on the QS's initial state), and keeping track of all the possible final energies of the electron becomes a rather difficult task. Furthermore, the size of the electron's Hilbert space will also depend on the initial electron's momentum distribution. All this indicates that trying to define the electron's Hilbert space beforehand will lead to problems of gargantuan proportions, since in principle one has to account for a continuum of possible final electron momenta. This is where the strategy outlined above shines: By performing all the algebra on the reduced Hilbert space of the QS (In out case all the states up to the 4$^{th}$ excitation manifold of the cavity + QE), one can obtain all the possible processes taking place within the QS, and find the associated effect these have on an arbitrary electron wavefunction. This allows us to effectively tackle the infinite-dimensional electron's Hilbert space for arbitrary initial states for the electron and QS and get analytical expressions for the different observables.\\

In order to computationally implement this strategy, we exploited the structure of \Cref{eqs:Scatt_matt_general} and the symbolic engine of Matlab. From \Cref{eqs:Scatt_matt_general} we see that the general scattering matrix can be obtained by first writing the interaction hamiltonian in matrix form in a basis of the reduced quantum system, and then by exponentiating this matrix one obtains the different matrix elements of the Scattering matrix. To complete the scattering matrix one needs to multiply each entry of this matrix by the appropriate electron ladder operator dictated by energy conservation. To perform the last part, we defined a new class of objects, "$b_{obj}$". These objects inherit the algebra of the $\hat{b}_q$ operators and allow to describe the interaction of the electron with a general system, without much thought about the particular electronic wavefunction or different momentum exchanges that might take place. We now give an outline of how these are implemented. These objecs are defined by two kinds of parameters: the amplitude, $\alpha$, and momentum exchange, $q_0$, so that: 
\begin{gather*}
    \alpha \hat{b}_{q_0} \equiv b_{obj}(\alpha,q_0)
\end{gather*}
where $\alpha$ is any complex number describing the amplitude of the operator and $q_0$ is the momentum shift of the ladder operator. These objects follow the following properties: 
\begin{enumerate}
    \item Hermitian conjugation: $(\alpha \hat{b}_q)^\dagger= \alpha^* \hat{b}_q^\dagger=\alpha^* \hat{b}_{-q}\rightarrow \left(b_{obj}(\alpha,q_0) \right) ^\dagger\equiv b_{obj}(\alpha^*,-q_0)$.
    \item Commutativity of ladder operators: $(\alpha \hat{b}_{q_0})(\beta \hat{b}_{q_1})=\alpha\beta \hat{b}_{q_0+q_1}\rightarrow b_{obj}(\alpha,q_0)b_{obj}(\beta,q_1)\equiv b_{obj}(\alpha\beta,q_0+q_1)$.
    \item Addition of several ladder operators: $b_{obj}(\alpha,q_0)+b_{obj}(\beta,q_1)\equiv b_{obj}([\alpha,\beta],[q_0,q_1])$.
    \item Distributive property: $b_{obj}(\gamma,q_2)b_{obj}([\alpha,\beta],[q_0,q_1]) \equiv b_{obj}([\alpha\gamma,\gamma\beta],[q_0+q_2,q_1+q_2])$.
\end{enumerate}
With these rules and performing the algebra on the reduced QS, in the end we have one of these objects with a list of the different momentum exchanges taking place in the system, and a list of the amplitudes corresponding to them.

\bibliography{bib_electrons}
